\documentclass[epj]{svjour}
\usepackage{graphics,wrapfig}
\usepackage[square,numbers,sort&compress]{natbib}
\usepackage{amsmath,amssymb}

\newcommand{\muin}[1]  {{\vec{\mathsf{#1}}}}
\newcommand{\pvec}[2]  {\begin{bmatrix}#1 \\ #2\end{bmatrix}}

\begin{document}
\title{\boldmath $N$-body simulations of gravitational dynamics}
\author{Walter Dehnen\inst{1} \and Justin I.\ Read\inst{1,2}}
\institute{Department of Physics \& Astronomy, University of Leicester,
  Leicester LE17RH, United Kingdom \and Institute for Astronomy, Department of
  Physics, ETH Z\"urich, Wolfgang-Pauli-Strasse 16, CH-8093 Z\"urich,
  Switzerland.}
\date{Received:  / Revised version: }
\abstract{We describe the astrophysical and numerical basis of $N$-body
  simulations, both of collisional stellar systems (dense star clusters and
  galactic centres) and collisionless stellar dynamics (galaxies and large-scale
  structure). We explain and discuss the state-of-the-art algorithms used for
  these quite different regimes, attempt to give a fair critique, and point out
  possible directions of future improvement and development. We briefly touch
  upon the history of $N$-body simulations and their most important results.
  \PACS{
    {98.10.+z}{Stellar dynamics and kinematics} \and
    {95.75.Pq}{Mathematical procedures and computer techniques} \and
    {02.60.Cb}{Numerical simulation; solution of equations}
  }
}
\maketitle

\section{Introduction}
\label{sec:intro}
Of the four fundamental forces, gravity is by far the weakest. Yet on large
distances it dominates all other interactions owing to the fact that it is
always attractive. Most gravitational systems are well approximated by an
ensemble of point masses moving under their mutual gravitational attraction and
range from planetary systems (such as our own) to star clusters, galaxies,
galaxy clusters and the universe as a whole.

Systems dominated by long-range forces such as gravity are not well treatable by
statistical mechanical methods: energy is not extensive, the canonical and
micro-canonical ensembles do not exist, and heat capacity is negative
\citep{BinneyTremaine2008}. Moreover, gravitational encounters are inefficient
for re-distributing kinetic energy, such that many such encounters are required
for relaxation, i.e.\ equipartition of kinetic energy. Gravitational systems,
where this process is potentially important over their lifetime are called
`collisional' as opposed to `collisionless' stellar systems\footnote{We shall
  use the term `stellar system' for a system idealised to consist of gravitating
  point masses, whether these are stars, planets, dark-matter particles, or
  whatever.}. The timescale over which this so-called `two-body relaxation' is
important is roughly \citep{BinneyTremaine2008}
\begin{equation} \label{eq:t:relax}
  t_{\mathrm{relax}} \simeq \frac{N}{8 \ln \Lambda}\,t_{\mathrm{dyn}}
\end{equation}
where $N$ is the number of particles; $t_{\mathrm{dyn}}$ is the dynamical
time\footnote{The dynamical or crossing time is a characteristic orbital time
  scale: the time required for a significant fraction of an orbit. A useful
  approximation, valid for orbits in inhomogeneous stellar systems, is
  $t_{\mathrm{dyn}}\simeq(G\bar{\rho})^{-1/2}$ with $\bar{\rho}$ the mean
  density interior to the particles current radius
  \citep{BinneyTremaine2008}. Note that the complete orbital period is longer by
  a factor $\sim3$.}; $\ln \Lambda = \ln (b_{\max}/b_{\min})$ is called the
\emph{Coulomb Logarithm}; and $b_{\max}$ and $b_{\min}$ are the maximum and
minimum impact parameters for the system, respectively\footnote{The minimum
  impact parameter is well defined for a system containing equal mass point
  particles: $b_{\min} = 2Gm/\sigma^2$, where $\sigma$ is the velocity
  dispersion of the background particles \citep{1976MNRAS.174...19W}. The
  maximum impact parameter is related in some way to the extent of the system,
  but is less well defined. Fortunately, even an order of magnitude uncertainty
  in $b_{\max}$ has only a small effect since it appears inside a
  logarithm. Notice that the form of the Coulomb Logarithm has a profound
  implication: each octave of $b$ contributes equally to $\ln\Lambda$ such that
  relaxation is driven primarily by weak \emph{long-range} interactions with
  $b\gg b_{\min}$.}. Collisional systems usually have a high dynamic age
($t_{\mathrm{dyn}}$ short compared to their lifetime) and high density, and
include globular star clusters and galactic centres. The majority of stellar
systems, however, are collisionless.

\begin{wrapfigure}{r}{82mm}
  \begin{center}
    \vspace{-8mm}
    \resizebox{70mm}{!}{\includegraphics{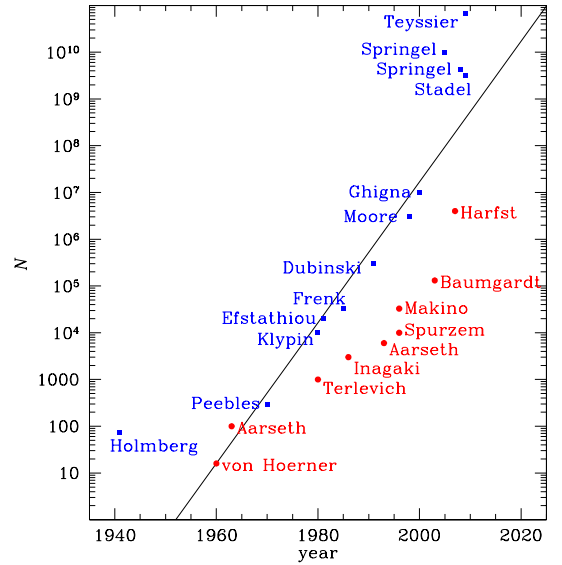}}
    \caption{\small\label{fig:nbodymoore}
      The increase in particle number over the past 50 years for selected
      collisional (red, \cite{1960ZA.....50..184V, Aarseth1963,
        1980IAUS...85..165T, 1986PASJ...38..853I, AarsethHeggie1993,
        1996MNRAS.282...19S, 1996ApJ...471..796M, 2003MNRAS.340..227B,
        HarfstEtAl2007} taken from \citep{2010NewAR..54..163H}) and
      collisionless (blue, \cite{1941ApJ....94..385H, Peebles1970,
        1981MNRAS.194..503E, 1985Natur.317..595F, DubinskiCarlberg1991,
        MooreEtal1998, 2000ApJ...544..616G, SpringelEtAl2005,
        2008MNRAS.391.1685S, TeyssierEtAl2009, 2009MNRAS.398L..21S}) $N$-body
      simulations. The line shows the scaling ${N{\,=\,}N_0
        2^{({\mathrm{year}}-y_0)/2}}$ expected from Moore's law if the
      costs scale $\propto N$.
    }
    \vspace{-7mm}
  \end{center}
\end{wrapfigure}

The numerical simulation of $N$-body systems has a long history, dating back to
the very first light-bulb experiments of \citet{1941ApJ....94..385H} in
1941. The first computer calculations were performed by
\citet{1960ZA.....50..184V} in 1960 who reached $N=16$, paving the way for the
pioneering work of \citet{Aarseth1963} in 1963 with $N=100$. Since these early
works, $N$ has nearly doubled every two years in accordance with Moore's law
(see Fig.~\ref{fig:nbodymoore}). The latest collisional $N$-body calculations
have reached over $10^6$ particles \citep{HarfstEtAl2007}, while collisionless
calculations can now reach more than $10^9$ particles \citep{SpringelEtAl2005,
  TeyssierEtAl2009, 2009MNRAS.398L..21S, IlievEtAl2011}.  This disparity
reflects the difference in complexity of these rather dissimilar $N$-body
problems. The significant increase in $N$ in the last decade was driven by the
usage of parallel computers.

In this review, we discuss the state-of-the art software algorithms and hardware
improvements that have driven this dramatic increase in $N$. We consider the
very different challenges posed by collisional (\S\ref{sec:collisional}) versus
collisionless (\S\ref{sec:less}) systems, and we attempt to give a fair critique
of the methods employed, pointing out where there is room for improvement, and
discussing some interesting future research directions. Our focus is primarily
on gravity; we do not consider the important role that the other fundamental
forces play\footnote{Many stellar systems contain significant amounts of gas,
  which in addition to gravity also interact electromagnetically. This gives
  rise to a large number of complicated effects from radiative cooling and the
  formation of stars (inside of which the strong and weak interactions become
  important, too), to active galactic nuclei and outflows driven by radiative
  heating. While for most gravitational systems these non-gravitational effects
  play an important role only for brief periods of their lifetime, their
  understanding and appropriate modelling is at the forefront of many
  contemporary challenges in astrophysics. This is, however, beyond the scope of
  this paper.}. Since our goal is to elucidate the numerics, we will only touch
upon the many interesting and important results that have come out of $N$-body
modelling over the past 50 years. We must therefore apologise in advance for all
that is missed out in this brief review. We do not have space to discuss
modelling gas physics and its many complications. Nor will we discuss the art of
setting up initial conditions for $N$-body simulations.

There are already several $N$-body reviews and books in the
literature. \citet{Aarseth2003} and \citet{HeggieHut2003} give excellent
reviews of the $N$-body problem, focusing mainly on collisional $N$-body
simulations. \citet{HockneyEastwood1988} cover many aspects of particle-based
simulations, focusing on collisionless applications. \citet{TrentiHut2008}
give a general overview of the $N$-body problem, covering both the collisional
and collisionless regimes. Our review takes a somewhat different approach to
these previous works. We attempt to review numerical techniques for both
collisional and collisionless $N$-body simulations, focusing on the very
latest techniques, and with a view to assessing interesting future research
directions. As much as possible, we present the key equations and describe the
methodology at a level where we hope the reader will be able to obtain a
relatively deep understanding of the algorithms and their hidden gremlins.

This paper is organised as follows. In \S\ref{sec:collisional}, we review
numerical methods for collisional $N$-body simulations: the astrophysical
(\S\ref{sec:coll:eqn}) and numerical foundations (\S\ref{sec:collforce}) with
special treatment of the time integration (\S\ref{sec:timestep}), recent
hardware-driven developments (\S\ref{sec:collisionalrecent}); and give a
critique of the current state of the art (\S\ref{sec:collcritique}), summarise
alternatives to $N$-body methods (\S\ref{sec:collalternative}), as well as
present a (brief and biased) overview over past and recent astrophysical results
with an outlook for the future (\S\ref{sec:coll:astr}). In \S\ref{sec:less}, we
review numerical methods for collisionless simulations: the astrophysical
(\S\ref{sec:less:eqn}) and numerical foundations (\S\ref{sec:less:num}), the
basis of cosmological $N$-body simulations (\S\ref{sec:cosmo}), force softening
(\S\ref{sec:soft}) and the various force solvers (\S\ref{sec:less:force}),
recent developments and challenges (\S\ref{sec:less:recent}), and a very brief
overview of astrophysical results (\S\ref{ref:less:astr}). \S\ref{sec:valid}
describes methods for validation of $N$-body simulations, and in
\S\ref{sec:conclusion} we present our conclusions and outlook for the future of
the field.

\section{N-body methods for collisional systems}
\label{sec:collisional}
Collisional systems are dynamically old such that $t_{\mathrm{dyn}}$ is short
compared to their age. This applies mainly to massive star clusters, and for
this reason we will mostly refer to the $N$-body particles in this section as
stars within a star cluster. Such clusters typically orbit deep within the
potential of a host galaxy -- like our own Milky Way -- such that their dynamics
is affected by the tidal field of their host.

Over many dynamical times, the accumulated effect of many small encounters
between stars significantly affects the evolution of collisional
systems. Relaxation-driven equipartition of energy causes heavier stars to sink
towards the centre, while low-mass stars are pushed to the outskirts, where they
are susceptible to being skimmed off by Galactic tides. Such relaxation
processes provide a mechanism for transporting specific energy outwards,
resulting in what is called the `gravothermal catastrophe': a dramatic increase
in the central density resulting in core collapse \citep{Antonov1962,
  LyndenBellWood1968, Spitzer1987, BaumgardtEtAl2003}.

In addition to relaxation -- which is largely driven by distant encounters --
short-range interactions can form bound particle pairs. Such binaries are
called `hard' if their binding energy exceeds the typical kinetic energy of
surrounding stars in the cluster. Close encounters with such field
stars result in a further hardening of hard binaries, while soft binaries
loose binding energy and become even softer (known as `Heggie's law'
\citep{Heggie1975, Hills1975}). Hard binary interactions are of great
importance, as they act as a source of kinetic energy
\citep{AarsethHills1972}, heating the core of the cluster (where most binaries
have sunk due to their larger mass), counter-acting the relaxation-driven
energy flux from the core to the outskirts, and thus prolonging the time until
core collapse \citep{FregeauEtAl2003, HeggieTrentiHut2006}. Core collapse
could perhaps even be reversed by binaries freshly formed in three-body
encounters (or two-body encounters involving strong stellar tides), a process
most likely to occur in the very high densities reached during core
collapse. The formation and evolution of hard binaries (and higher order
multiples of stars) presents a unique numerical challenge, because their short
dynamical times and extreme forces lead to very small integration time steps
\citep{Aarseth2003}; they require special treatment that we discuss in
\S\ref{sec:timestep}.

Another reason binary interactions are important astrophysically is that they
provide a route to forming close binary stars of a constitution and type
unlikely to form under ordinary star-formation conditions. `Blue Straggler'
stars, for example, may form in this way \citep{Bailyn1995,
  PeretsFabrycky2009, LanzoniEtAl2007}, as well as ultra compact X-ray
binaries, which are over-abundant in globular clusters \citep[and references
  therein]{King2011}.

\subsection{Equations governing collisional stellar systems}
\label{sec:coll:eqn}
The physics of collisional stellar systems is in principle quite simple: the
motion of $N$ point masses under their mutual gravitational attraction. This
is in fact a Hamiltonian system, with a Hamiltonian $H$ and equations of
motion
\begin{equation}
  H = \sum_i \frac{\vec{p}_i^2}{2m_i} - G \sum_i\sum_{j>i}
  \frac{m_i\,m_j}{|\vec{x}_i-\vec{x}_j|},\qquad
 {\vec{a}}_i = \frac{\dot{\vec{p}_i}}{m_i}
  = - \frac{1}{m_i} \frac{\partial H}{\partial\vec{x}_i}
  = -G\sum_{j\neq i}
    m_j\,\frac{\vec{x}_i-\vec{x}_j}{|\vec{x}_i-\vec{x}_j|^3}.
\label{eqn:direct}
\end{equation}
where $\vec{p}_i = m_i \dot{\vec{x}}_i$ is the momentum; $\vec{x}_i$ is the
position; and $\vec{a}_i = \ddot{\vec{x}}_i$ is the acceleration of particle
$i$. $N$-body simulations of collisional systems simply try to solve these
equations directly by brute force.

For problems involving massive central black holes, general relativistic (GR)
corrections to the force can become important. This is because, although we are
almost always in the weak field regime, if the black hole dominates the central
potential, then the potential will be close to Keplerian and super-resonant. GR
effects act to break resonance by inducing orbital precessions. Over many
dynamical times, or if stars orbit very close to the central black hole,
such effects can become important \citep{Aarseth2007, HarfstEtAl2008}. We will
not discuss such `post-Newtonian' corrections to the force further in this
review, but refer the interested reader to \citet{Aarseth2007} and references
therein.

\subsection{Numerics of collisional N-body simulations}
\label{sec:collforce}
While equations~(\ref{eqn:direct}) are conceptually straightforward, they are
anything but straightforward to solve numerically. One obvious problem is that
the computational cost for the force calculation for all particles grows as
$\mathrm{O}(N^2)$, implying that realistic simulations with $N\sim10^6$ (the
number of stars in massive star clusters) are challenging. Such high $N$ has
only recently been achieved for a small number of simulations using special
hardware chips operated in parallel \citep{HarfstEtAl2007} (see also
\S\ref{sec:collisionalrecent}). A second serious problem is the enormous range
of time scales, ranging from days for the periods of tight
binaries\footnote{\label{foot:4U1820-30} The shortest observed binary period
  is 11 minutes for the X-ray binary 4U\,1820-30 in the globular cluster
  NGC6624 \citep{MorganRemillardGarcia1988}, but such very short periods are
  not simulated in contemporary $N$-body simulations.} to millions of years
for the orbits of most stars, and $10^{10}$ years for the age of the whole
stellar system. To make things worse, the time step required to accurately
integrate the trajectory of any individual star can change considerably along
its orbit and abruptly during close encounters. Given this range of formidable
problems, it is not surprising that there are very few $N$-body codes which
can cope with them.

As already alluded to, the computation of the forces dominates the computational
cost of existing collisional $N$-body codes, all of which use the brute-force
direct-summation approach (i.e.\ a straightforward implementation of
equations~\ref{eqn:direct}). This is motivated by the need for an accurate force
computation to ensure correct modelling of both close and distant encounters.
Currently, several approaches are applied to reduce the frequency of force
computations for any individual star. First, the force is split into
contributions from near neighbours and the far field. While the former varies on
short time scales, it is quite cheap to compute, whereas the latter is expensive
to evaluate but much smoother in time. Thus, splitting the force into two
components like this allows us to reduce the need for the expensive computation
of the far-field force. This `Ahmad-Cohen' \citep{AhmadCohen1973} scheme
necessarily requires individual timesteps for each particle which we discuss in
\S\ref{sec:timestep}.

Second, the frequency for computing the acceleration $\vec{a}_i$ can be further
reduced by computing its time derivative, the jerk:
\begin{equation}
  \dot{\vec{a}}_i = -G\sum_{j\neq i}m_j
  \frac{\vec{x}_{i\!j}^2\dot{\vec{x}}_{i\!j}
    -3\vec{x}_{i\!j}(\vec{x}_{i\!j}\cdot\dot{\vec{x}}_{i\!j})}{|\vec{x}_{i\!j}|^5}
    \qquad\mathrm{with}\qquad\vec{x}_{i\!j}\equiv\vec{x}_i-\vec{x}_j,
\label{eqn:jerk}
\end{equation}
which is used to predict ${\vec{a}}_i$ into the future or, equivalently, to
employ a higher-order time integrator with a larger time step (see
\S\ref{sec:timestep}). Ever higher order schemes require ever higher derivatives
of $\vec{a}_i$ to obtain forward interpolations of the force
\citep{NitadoriMakino2008}.

\subsection{Time integration}\label{sec:timestep}
The accurate time integration of close encounters is the most difficult part
of collisional $N$-body methods, while for collisionless $N$-body methods
force softening (see \S\ref{sec:soft}) alleviates this problem
substantially. Here, we review the various time integration methods employed
in both types of $N$-body methods. Let us begin our considerations by the
simple `Euler method', which updates the position and velocity for a given
particle by \emph{timestep} $\Delta t$ via
\begin{subequations}
  \label{eqn:simpleeuler}
  \begin{eqnarray}
    \vec{x}(t + \Delta t) &=& \vec{x}(t) + \dot{\vec{x}}\, \Delta t
    \label{eqn:simpleeulerx} \\[0.5ex]
    \dot{\vec{x}}(t + \Delta t) &=& \dot{\vec{x}}(t) + \vec{a}(t)\,
    \Delta t.
  \label{eqn:simpleeulerv}
  \end{eqnarray}
\end{subequations}
While conceptually straightforward, this scheme performs very poorly in
practice. The Euler method is just a Taylor expansion to first order in $\Delta
t$ and the errors are proportional to $\Delta t^2$. We can significantly improve
on this at little additional computational cost either by increasing the
expansion order and thus the accuracy, or by integrating a `near-by' Hamiltonian
\emph{exactly} using a low-oder scheme. We now compare and contrast a popular
example of each type of approach: the second-order leapfrog integrator, which is
heavily used in collisionless $N$-body applications, and the fourth-order
Hermite scheme, which has become the integrator of choice for collisional
applications.

\subsubsection{The Leapfrog integrator}
\label{sec:leapfrog}
The leapfrog integrator is an example of a \emph{symplectic
  integrator}. Symplectic integrators exactly solve an approximate Hamiltonian.
As a consequence, the numerical time evolution is a canonical map and preserves
certain conserved quantities exactly, such as the total angular momentum, the
phase-space volume, and the Jacobi constants. The idea is to approximate the
Hamiltonian $H$ in equation~(\ref{eqn:direct}) with
\begin{equation}
  \tilde{H} = H + H_{\mathrm{err}}
  \label{eqn:Ha}
\end{equation}
where $H_{\mathrm{err}}$ is the error Hamiltonian. Provided that $\tilde{H}$ and
$H$ are time-invariant, the energy error is bounded at all times
\citep{Yoshida1993}\footnote{A symplectic integrator which obtains zero energy
  error is exact.}. The goal now is to find $\tilde{H}$ that can be solved
\emph{exactly} by simple numerical means and minimises
$H_{\mathrm{err}}$. Defining the combined phase-space coordinates $\vec{w} =
(\vec{x},\vec{p})$ we can re-write Hamilton's equations as:
\begin{equation}
   \dot{\vec{w}} = \mathcal{H} \vec{w},
  \label{eqn:hamilfunny}
\end{equation}
where $\mathcal{H}\equiv\left\{\cdot, H\right\}$ (with the Poisson bracket
$\{A,B\}\equiv \partial_{\vec{x}}A{\cdot}\partial_{\vec{p}}B-
\partial_{\vec{x}}B{\cdot}\partial_{\vec{p}}A$) is an \emph{operator} acting
on $\vec{w}$. Equation~(\ref{eqn:hamilfunny}) has the formal solution
\begin{equation}
  \vec{w}(t+\Delta t) = \mathrm{e}^{\Delta t\,\mathcal{H}}\,\vec{w}(t),
\end{equation}
where we can think of the operator $\mathrm{e}^{\Delta t\,\mathcal{H}}$ as a
symplectic map from $t$ to $t+\Delta t$. This operator can be split, in an
approximate sense, into a succession of discrete but symplectic steps, each of
which can be \emph{exactly} integrated. The most common choice is to separate
out the kinetic and potential energies, $H=T(\vec{p}) + V(\vec{x})$, such that
we can split
\begin{equation} \label{eqn:opsplit}
  \mathrm{e}^{\Delta t \,\mathcal{H}} = 
  \mathrm{e}^{\Delta t \,(\mathcal{T}+\mathcal{V})} \simeq
  \mathrm{e}^{\Delta t \,\mathcal{V}}\,
  \mathrm{e}^{\Delta t \,\mathcal{T}} = 
  \mathrm{e}^{\Delta t \,\tilde{\mathcal{H}}}.
\end{equation}
Because the differential operators $\mathcal{T}\equiv\{\cdot,T\}$ and
$\mathcal{V}\equiv\{\cdot,V\}$ are \emph{non-commutative}, the central relation
in equation~(\ref{eqn:opsplit}) is only approximately true. This operator
splitting is extremely useful, because, while equation~(\ref{eqn:hamilfunny})
has in general no simple solution, the equivalent equations for each of our new
operators do:
\begin{equation} \label{eqn:kick:drift}
  \mathrm{e}^{\Delta t \,\mathcal{T}} \pvec{\vec{x}}{\vec{p}} = 
  \pvec{\vec{x}+\Delta t\,\vec{p}}{\vec{p}}
  \qquad\text{and}\qquad
  \mathrm{e}^{\Delta t \,\mathcal{V}} \pvec{\vec{x}}{\vec{p}} = 
  \pvec{\vec{x}}{\vec{p}-\Delta t\,\vec{\nabla}V(\vec{x})}.
\end{equation}
These operations are also known as \emph{drift} and \emph{kick} operations,
because they only change either the positions (drift) or velocities
(kick). Note that the drift step in (\ref{eqn:opsplit}) is identical to the
simple Euler method~(\ref{eqn:simpleeulerx}), while its kick step is
\emph{not} identical, because the acceleration is calculated using the drifted
rather than the initial positions. The integrator that applies a drift
followed by a kick (equation \ref{eqn:opsplit}) is called \emph{modified}
Euler scheme and is symplectic.

It is clear from the similarity between the simple and modified Euler schemes
that both are only first order accurate. We can do better by concatenating many
appropriately weighted kick and drift steps:
\begin{equation} \label{eqn:hhat}
  \mathrm{e}^{\Delta t \,\tilde{\mathcal{H}}} = \prod_i^N
  \mathrm{e}^{a_i\Delta t\mathcal{V}}\,\mathrm{e}^{b_i\Delta t\mathcal{T}}
  = \mathrm{e}^{\Delta t \,\mathcal{H} + \mathcal{O}(\Delta t^{n+1})}
\end{equation}
with coefficients $a_i$ and $b_i$ chosen to obtain the required order of
accuracy $n$. From equation~(\ref{eqn:hhat}) we see that: (i) the approximate
Hamiltonian $\tilde{H}$ is solved exactly by the successive application of the
kick and drift operations, and (ii) $\tilde{H}$ approaches $H$ in the limits
$\Delta t\to0$ or $n\to\infty$. At second order ($n=2$), and choosing
coefficients that minimise the error, we derive the \emph{leapfrog} integrator:
\begin{equation}
  \label{eq:leapfrog}
  \mathrm{e}^{\Delta t \,\mathcal{H} + \mathcal{O}(\Delta t^3)}=
  \mathrm{e}^{\frac{1}{2}\Delta t\,\mathcal{V}} \mathrm{e}^{\Delta
    t\,\mathcal{T}} \mathrm{e}^{\frac{1}{2}\Delta t\,\mathcal{V}}.
\end{equation}
Applying equations~(\ref{eqn:kick:drift}), this becomes (subscripts 0 and 1
refer to times $t$ and $t+\Delta t$, respectively):
\begin{subequations}
  \label{eqn:kdk:a}
  \begin{eqnarray}
    \label{eqn:kick1}
    \dot{\vec{x}}' &=& \dot{\vec{x}}_0 +
    \tfrac{1}{2}\vec{a}_0\Delta t \\
    \label{eqn:drift}
    \vec{x}_1 &=& \vec{x}_0 +\phantom{\tfrac{1}{2}}
    \dot{\vec{x}}'\Delta t \\
    \label{eqn:kick2}
    \dot{\vec{x}}_1&=&\dot{\vec{x}}^\prime\, + 
    \tfrac{1}{2}\vec{a}_1\Delta t
  \end{eqnarray}
\end{subequations}
where $\vec{a}_0=-\vec{\nabla}V(\vec{x}_0)$ and
$\vec{a}_1=-\vec{\nabla}V(\vec{x}_1)$, while the intermediate velocity
$\dot{\vec{x}}'$ serves only as an auxiliary quantity. Combining
equations~(\ref{eqn:kdk:a}) we find the familiar Taylor expansions of the
positions and velocities to second order\footnote{This is the kick-drift-kick
  (KDK) leapfrog. An alternative is the drift-kick-drift (DKD) version. In
  practice, the KDK is preferable, because acceleration and potential are known
  at the second-order accurate positions (not at an auxiliary intermediate
  position as for the DKD), facilitating their usage in time-step
  control. Moreover, with the block-step scheme (see \S\ref{sec:step:choice} and
  Fig.~\ref{fig:timestep}) the KDK results in synchronised force computations
  for all active particles.}:
\begin{subequations}
  \label{eqn:kdk:b}
  \begin{eqnarray}
    \label{eqn:leapfrogimp1}
    \vec{x}_1 &=& \vec{x}_0 +
    \dot{\vec{x}}_0\Delta t + \tfrac{1}{2} \vec{a}_0\Delta t^2,\\
    \label{eqn:leapfrogimp2}
    \dot{\vec{x}}_1 &=& \dot{\vec{x}}_0 +
    \tfrac{1}{2} (\vec{a}_0+\vec{a}_1)\Delta t.
  \end{eqnarray}
\end{subequations}

In principle, one may combine as many kick and drift operations as desired to
raise the order of the scheme. However, it is impossible to go beyond second
order without having at least one $a_i$ and one $b_i$ coefficient in
equation~(\ref{eqn:hhat}) be negative \citep[][see also
  \citenum{LeimkuhlerReich2005}]{Sheng1989,Suzuki1991}. This involves some
\emph{backwards} integration and is problematic when using variable
timesteps---especially if time symmetry is required\footnote{Recently,
  \citet{ChinChen2005} have constructed fourth-order symplectic integrators
  which require only forwards integration. To achieve this, rather than
  eliminate all the errors by appropriate choice of the coefficients $a_i$ and
  $b_i$, they \emph{integrate} one of the error terms thus avoiding any backward
  step. Their method requires just two force and one force gradient evaluation
  per time step. It has not yet found application in $N$-body dynamics, but
  could be a very promising avenue for future research. \label{foot:4symp}}.

As mentioned previously, variable individual timesteps are a requirement for
most $N$-body applications. Unfortunately, once we allow $\Delta t$ to vary in
space and time, the symplectic nature of the leapfrog is broken. But, we can do
almost as well as symplectic by ensuring that the scheme remains \emph{time
  symmetric} \citep{QuinlanTremaine1990}. One route to time symmetry is to
solve the \emph{implicit} leapfrog equations~(\ref{eqn:kdk:b}) using a
symmetrised timestep:
\begin{equation}
  \overline{\Delta t} = \tfrac{1}{2} \big[T(\vec{x}_{0},
    \dot{\vec{x}}_{0}, \vec{a}_{0} ...) + T(\vec{x}_{1},
    \dot{\vec{x}}_{1}, \vec{a}_{1} ...) \big],
\end{equation}
where the function $T$ generates the step size depending on the position,
velocity, acceleration etc. (we will discuss possible functional forms in
\S\ref{sec:step:choice}). Since $\overline{\Delta t}$ is a function of the force
evaluated at $t+\Delta t$, which in turn is a function of $\overline{\Delta t}$,
such a scheme is implicit and in general requires an iterative solution
\citep{HutMakinoMcMillan1995}.  Since each iteration involves another expensive
force evaluation, implicit schemes are not used in practice. Fortunately, there
are explicit and time-symmetric methods for adapting the time step
\citep{HolderLeimkuhlerReich1999}, for example
\begin{equation}
  \Delta t_{\mathrm{old}}\, \Delta t_{\mathrm{new}} = T(\vec{x},
  \dot{\vec{x}}, \vec{a} ... )^2,
  \label{eqn:symtimestep}
\end{equation}
where $\Delta t_{\mathrm{old}}$ and $\Delta t_{\mathrm{new}}$ are the time steps
used to evolve \emph{to} and \emph{from} the arguments of $T$. Clearly,
equation~(\ref{eqn:symtimestep}) is time symmetric by construction and requires
no iteration to solve.

\begin{figure}
  \begin{center}
    \hfil
    \resizebox{50mm}{!}{\includegraphics{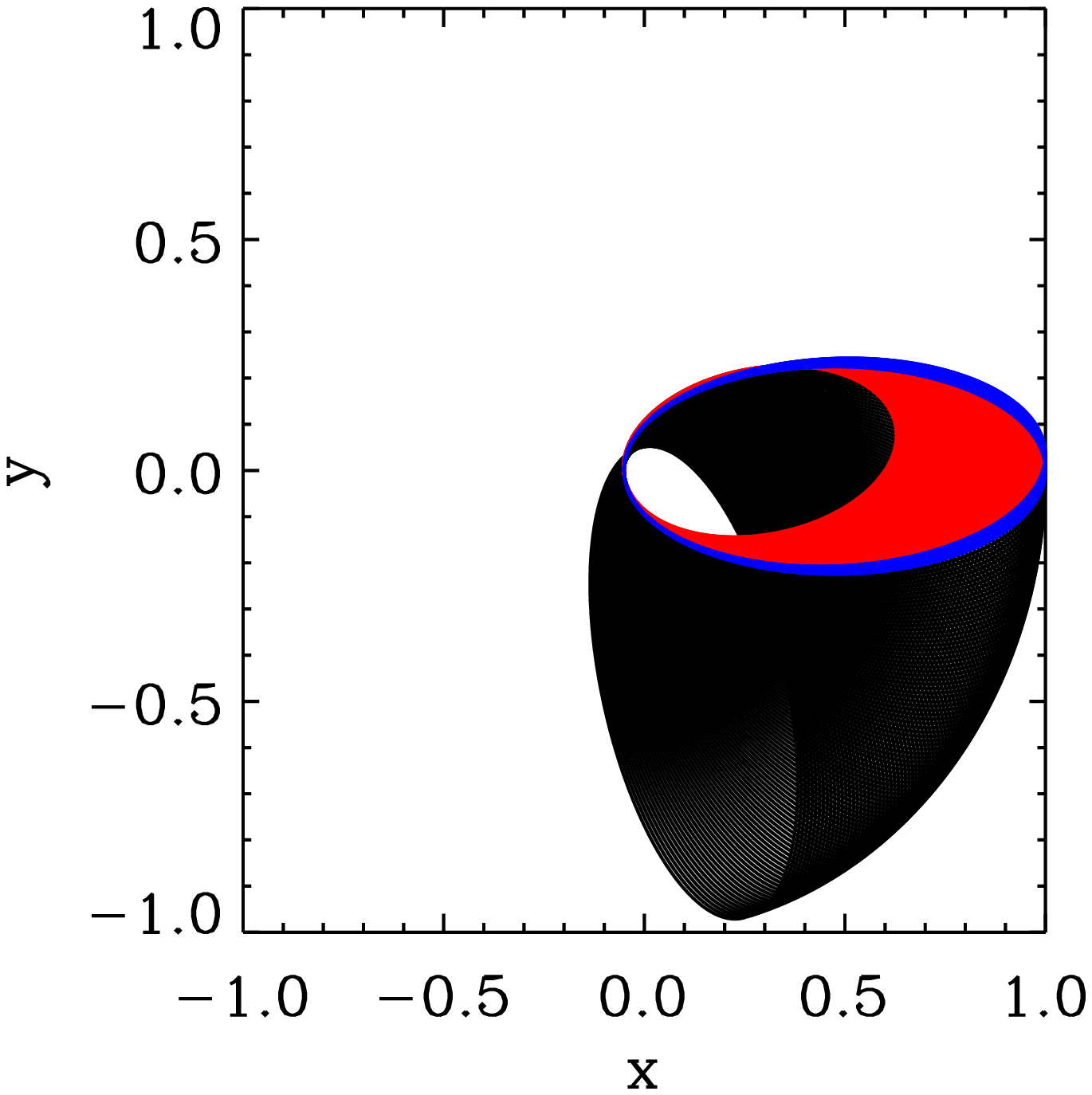}}\hfil
    \resizebox{50mm}{!}{\includegraphics{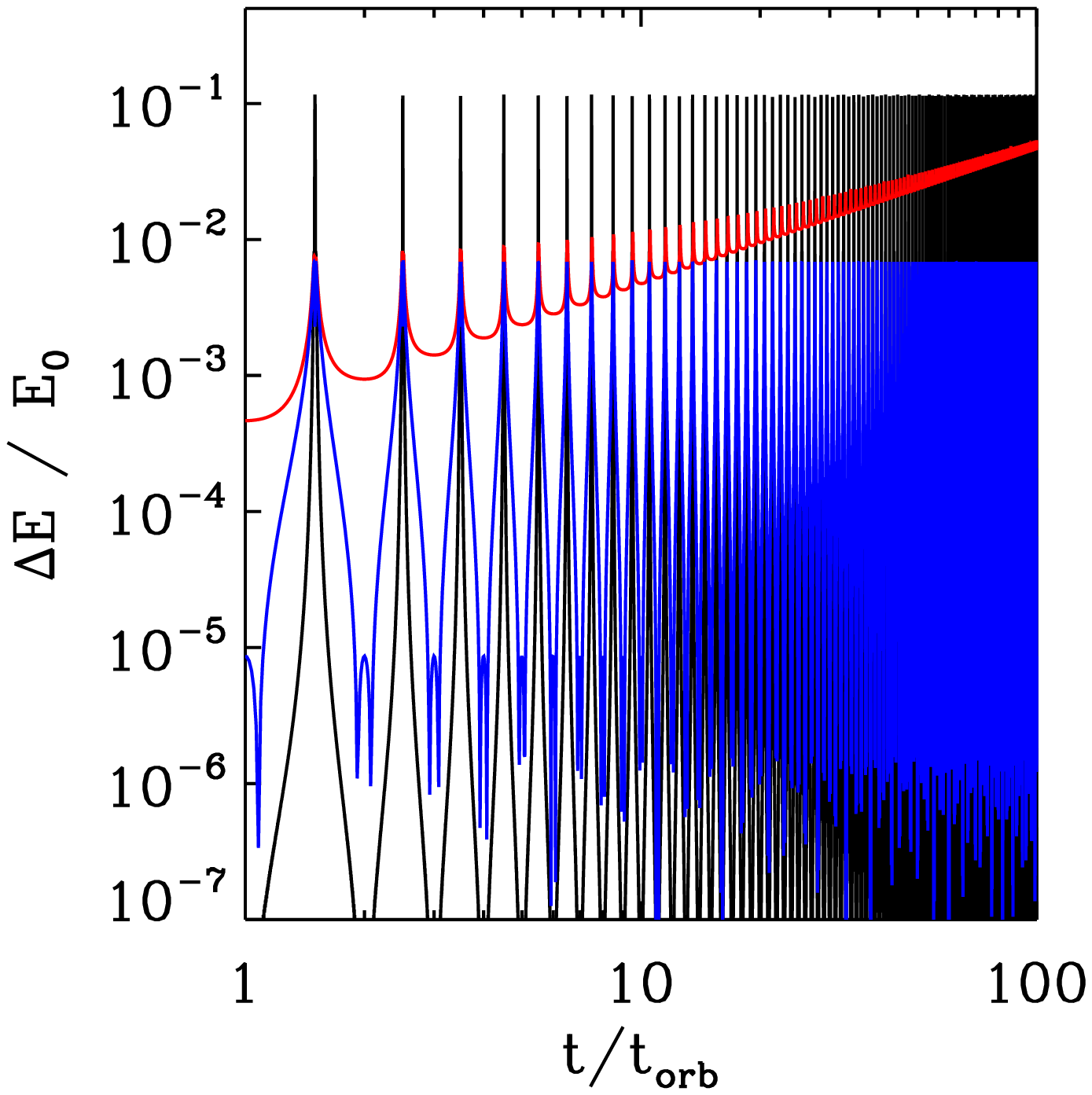}}\hfil
    \resizebox{50mm}{!}{\includegraphics{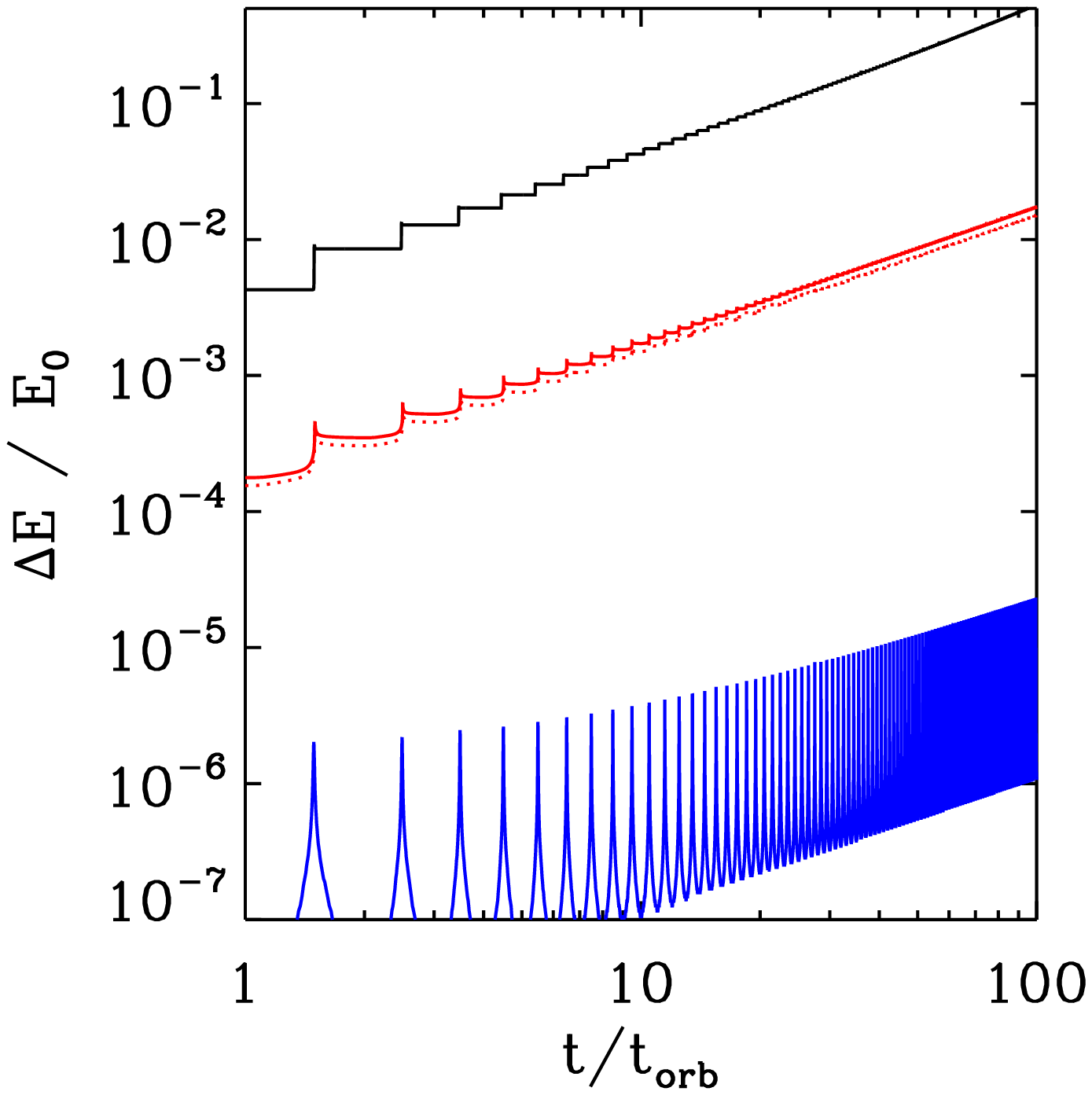}}\hfil
    \caption{\small\label{fig:force_schematic}
      \textbf{Left} Comparison of the leapfrog integrator (black); a 4th order
      Hermite scheme (red); and time symmetric leapfrog using variable timesteps
      (blue) for the integration of an elliptic ($e=0.9$) Kepler orbit over 100
      periods. In the first two cases a fixed timestep of $0.001$ of the period
      was used; in the latter case we use the timestep criteria in
      equation~(\ref{eqn:leapfrogstep}).
      \textbf{Middle} Fractional change in energy for the Kepler problem for the
      leapfrog integrator with fixed timesteps (black), variable timesteps
      (equation \ref{eqn:leapfrogstep}; red), and symmetric variable timesteps
      (blue).
      \textbf{Right} Fractional change in energy for the Kepler problem for the
      4th order Hermite integrator with fixed timesteps (black) variable
      timesteps (equation \ref{eqn:aarsethstep}; red), and variable timesteps
      (equation \ref{eqn:justinstep}; red dotted). The blue curve shows the
      energy error for the same Kepler orbit calculated using the K-S
      regularised equations of motion (see text for details). All calculations
      with variable timesteps were run at the same computational cost ($\sim250$
      force and jerk evaluations per orbit, which is about a quarter of the cost
      of the fixed-timestep calculations).
    }
  \end{center}
\end{figure}
In the middle panel of Fig.~\ref{fig:force_schematic}, we compare energy
conservation for the leapfrog using fixed timesteps, variable timesteps, and
time symmetric variable timesteps of equation~(\ref{eqn:symtimestep}) for a
Kepler orbit with eccentricity $e = 0.9$. With fixed timesteps (black) the
energy fluctuates on an orbital time scale, but is perfectly conserved in the
long term; with variable timesteps, manifest energy conservation is lost
(red); while with the time symmetric variable time step scheme, we recover
excellent energy conservation (blue). The time symmetric variable time step
leapfrog used about a quarter of the force calculations required for the
fixed-step integration while giving over an order of magnitude better energy
conservation. This is why variable timesteps are an essential ingredient in
modern $N$-body calculations.

\subsubsection{Hermite integrators}
\label{sec:hermite}
The leapfrog integrator is a popular with collisionless $N$-body applications
because of its simplicity, manifest energy conservation, and stability. However,
its integration errors on short time scales make it less useful for studying
collisional systems, where one must correctly track chaotic close encounters,
and such errors would rapidly ruin the integration. Shrinking the timestep
helps, but as the leapfrog is only a second-order scheme, the step sizes
required become prohibitively small. This motivates considering higher-order
non-symplectic integrators for collisional applications.

The current state of the art are fourth-order non-symplectic integrators, so
called \emph{Hermite} schemes (for higher orders see
\citep{NitadoriMakino2008}). From the Taylor expansion of the position and its
time derivatives at time $t+\Delta t$
\begin{subequations}
  \label{eqn:hermite:taylor}
  \begin{eqnarray}
  \label{eqn:hermite:taylor:x}
  \vec{x}_{1} &=&
  \vec{x}_{0} + \dot{\vec{x}}_0 \Delta t +
  \tfrac{1}{2}\vec{a}_{0} \Delta t^2 +
  \tfrac{1}{6}\dot{\vec{a}}_{0} \Delta t^3 +
  \tfrac{1}{24}\ddot{\vec{a}}_{0} \Delta t^4,
  \\[0.5ex]
  \label{eqn:hermite:taylor:v}
  \dot{\vec{x}}_{1} &=&
  \dot{\vec{x}}_{0} + \vec{a}_{0} \Delta t +
  \tfrac{1}{2}\dot{\vec{a}}_{0} \Delta t^2 +
  \tfrac{1}{6}\ddot{\vec{a}}_{0} \Delta t^3 +
  \tfrac{1}{24}\dddot{\vec{a}}_{0} \Delta t^4,
  \\[0.5ex]
  \label{eqn:hermite:taylor:a}
  \vec{a}_{1} &=&
  \vec{a}_{0}+\dot{\vec{a}}_0 \Delta t +
  \tfrac{1}{2}\ddot{\vec{a}}_{0} \Delta t^2 +
  \tfrac{1}{6}\dddot{\vec{a}}_{0} \Delta t^3,
  \\[0.5ex]
  \label{eqn:hermite:taylor:j}
  \dot{\vec{a}}_{1} &=&
  \dot{\vec{a}}_{0}+\ddot{\vec{a}}_0 \Delta t +
  \tfrac{1}{2}\dddot{\vec{a}}_{0} \Delta t^2,
  \end{eqnarray}
\end{subequations}
we can eliminate $\ddot{\vec{a}}_0$ and $\dddot{\vec{a}}_0$ to obtain
\begin{subequations} \label{eq:hermite:scheme}
  \begin{eqnarray}
    \vec{x}_{1} &=& \vec{x}_{0} + 
    \tfrac{1}{2}  (\dot{\vec{x}}_{1} + \dot{\vec{x}}_{0}) \Delta t +
    \tfrac{1}{12} (\vec{a}_{0}-\vec{a}_{1}) \Delta t^2
    \;+\mathcal{O}(\Delta t^5),
    \\[0.5ex]
    \dot{\vec{x}}_{1} &=& \dot{\vec{x}}_{0} +
    \tfrac{1}{2} ({\vec{a}}_{1} + \vec{a}_{0}) \Delta t + 
    \tfrac{1}{12}( \dot{\vec{a}}_{0} - \dot{\vec{a}}_{1}) \Delta t^2
    \;+\mathcal{O}(\Delta t^5).
  \end{eqnarray}
\end{subequations}
These equations are not only fourth-order accurate but also time symmetric
(though not symplectic) and will therefore give excellent energy conservation.
The only snag is their circularity: in order to obtain $\vec{x}_1$ and
$\dot{\vec{x}}_1$ we need to know the acceleration $\vec{a}$ and jerk
$\dot{\vec{a}}$ not only at time $t$ (where we can readily compute them from
equations~\ref{eqn:direct} and \ref{eqn:jerk}) but also at time $t+\Delta t$,
when they in turn depend on $\vec{x}_1$ and $\dot{\vec{x}}_1$. In order words,
equations~(\ref{eq:hermite:scheme}) define an \emph{implicit} scheme. In
practice, this difficulty is side-stepped by first \emph{predicting} positions
and velocities
\begin{subequations}
  \label{eqn:predictor}
  \begin{eqnarray}
    \label{eqn:predictor1}
    \vec{x}_{p} &=& \vec{x}_{0} + \dot{\vec{x}}_{0} \Delta t +
    \tfrac{1}{2}{\vec{a}}_{0} \Delta t^2 + \tfrac{1}{6}\dot{\vec{a}}_{0}
    \Delta t^3,
    \\ \label{eqn:predictor2}
    \dot{\vec{x}}_{p} &=& \dot{\vec{x}}_{0} +{\vec{a}}_{0} \Delta t +
    \tfrac{1}{2}\dot{\vec{a}}_{0} \Delta t^2;
  \end{eqnarray}
\end{subequations}
then \emph{estimating} acceleration $\vec{a}_1$ and jerk $\dot{\vec{a}}_1$ using
equations~(\ref{eqn:direct}) and (\ref{eqn:jerk}) with the predicted positions
and velocities; and finally obtaining the \emph{corrected} $\vec{x}_1$ and
$\dot{\vec{x}}_1$ from equations~(\ref{eq:hermite:scheme}). A single iteration
of this method is is called a Predict-Evaluate-Correct (PEC) scheme; further
iterations are denoted P(EC)$^n$, with $n$ the number of iterations
\citep{KokuboYoshinagaMakino1998}. In the limit $n\to\infty$, we converge on the
implicit Hermite solution~(\ref{eq:hermite:scheme}). In practice, implicit
integration schemes are not employed because of the numerical cost of
calculating the acceleration and jerk over several iterations. However, unlike
the implicit Hermite scheme, the explicit PEC scheme is not time symmetric.

A comparison of various flavours of the 4th order Hermite integrator are given
in Figure \ref{fig:force_schematic}, for the integration of an elliptical
Kepler orbit with $e=0.9$ over 100 orbits. Notice that the leapfrog integrator
with fixed timestep conserves energy exactly (in the long-term), but that the
peak of the oscillations is initially $\sim$ two orders of magnitude worse
than for the 4th order Hermite scheme with fixed steps (compare black lines in
middle and right panels). Over time, the energy losses accumulate for the
Hermite integrator, causing the apocentre of the orbit to decay, but the
orbital precession (both are numerical errors) is significantly less than for
the leapfrog (compare black and red orbits in the left panel). It is the
orbital stability and excellent energy accuracy that have made Hermite
integrators popular for use in collisional $N$-body problems.

\subsubsection{The choice of time-step}
\label{sec:step:choice}
Given the enormous dynamic range in time involved in collisional $N$-body
problems (ranging from days to giga-years), it has become essential to use
variable timestep schemes \citep{Aarseth2003}. Early schemes used an
individual time step for each particle. However, it is better to arrange the
particles in a hierarchy of timesteps organised in powers of two, with
reference to a `base step' $\Delta t_0$ \citep{Makino1991:B}:
\begin{equation}
  \Delta t_n = \Delta t_0 / 2^{n}
\end{equation}
\begin{wrapfigure}{r}{70mm}
  \begin{center}
    \vspace{-3mm}
    \resizebox{55mm}{!}{\includegraphics{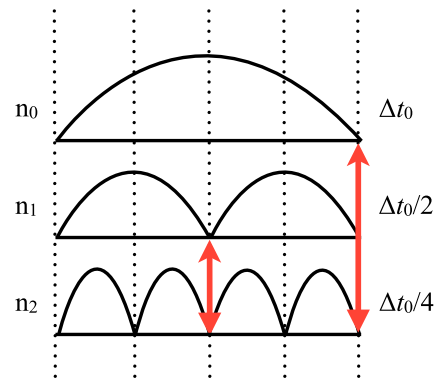}}
    \vspace{5mm}
    \caption{\small\label{fig:timestep}
      Schematic illustration of a block time-stepping scheme. Particles
      are organised on timesteps in a hierarchy of powers of two relative to a
      base time $\Delta t_0$. The time step level, denoted $n_{0,1,2 ...}$ is
      called the timestep {\it rung}. Particles can move up and down rungs at
      synchronisation points marked by the red arrows.
    }
    \vspace{-4mm}
  \end{center}
\end{wrapfigure}
for a given timestep \emph{rung} $n$. Particles can then move between rungs at
synchronisation points as shown in Fig.~\ref{fig:timestep}. This
\emph{block-step} scheme leads to significant efficiency savings because
particles on the same rung are evolved simultaneously. However, time symmetry
with block stepping presents some challenges \citep{MakinoEtAl2006}. A
key problem is that, in principle, particles can move to lower timestep rungs
whenever they like, but they may only move to higher rungs at synchronisation
points where the end of the smaller step overlaps with the end step of a
higher rung (see Fig.~\ref{fig:timestep}). This leads to an asymmetry in the
timesteps, even if some discrete form of equation~(\ref{eqn:symtimestep}) is
used. \citet{MakinoEtAl2006} show that it is possible to construct a
near-time symmetric block time step scheme, provided some iteration is allowed
in determining the time step. Whether a non-iterative scheme is possible
remains to be seen.

We now need some criteria to decide which rung a particle should be placed
on. For low-order integrators like the leapfrog, we have only the acceleration
to play with. In this case, a possible timestep criterion can be found by
analogy with the Kepler problem:
\begin{equation}
  \label{eqn:leapfrogstep}
  \Delta t_i = \eta \sqrt{|\Phi_i|}/|\vec{a}_i|,
\end{equation}
where $\Phi_i$ is the gravitational potential of particle $i$, and $\eta$ is a
dimensionless accuracy parameter. Substituting $\Phi_i = GM/r_i$ and
$|\vec{a}_i| = GM/r_i^2$, valid for a particle at radius $r_i$ orbiting a
point mass $M$, we see that this gives $\Delta t_i=\eta \sqrt{r_i^3/GM}$,
i.e.\ exactly proportional to the dynamical time. However, a timestep criteria
that depends on the potential is worrisome since the transformation
$\Phi\to\Phi+\mathrm{const.}$ has no dynamical effect, but would alter the
timesteps. In applications like cosmological $N$-body simulations, where the
local potential has significant external contributions, simulators have
typically employed:
\begin{equation}
  \label{eqn:cosmostep}
  \Delta t_i = \eta \sqrt{\epsilon/|\vec{a}_i|}
\end{equation}
and similar, where $\epsilon$ is the force softening length (see
\S\ref{sec:less}). Equation~(\ref{eqn:cosmostep}) is really only
defined on dimensional grounds: it creates a quantity with dimensions of time
from a local length scale---the force softening---and the local
acceleration. It is clear that this time step criteria would be of no use for,
say, the Kepler problem, where it will lead to too small steps at large radii,
and too large steps at small radii.

In a recent paper, \citet{ZempEtAl2007} have attempted to solve the above
conundrum by trying to determine what a particle is orbiting about. If this is
known, then the dynamical time itself makes for a natural timestep criteria:
\begin{equation}
  \label{eqn:idealstep}
  \Delta t_i = \eta \sqrt{r_i^3/GM(r_i)},
\end{equation}
where $M(r_i)$ is the mass enclosed within the particle's orbit from some
`attractor' at distance $r_i$. \citeauthor{ZempEtAl2007} attempted to define a
$M(r)$ based on information taken from a gravitational tree structure (see
\S\ref{sec:less}). Such ideas lend themselves naturally to collisionless
simulations, where a tree is often readily available as a by-product of the
force calculation (see \S\ref{sec:tree:fmm}). But it remains to be seen if
such a timestep criteria can be competitive for collisional $N$-body
applications. Unlike many collisionless applications, in collisional $N$-body
applications, it is often well-defined from the outset what particles are
orbiting about---at least until close interactions occur when case special
treatment is required anyway. In addition, the higher-order integrators
typically employed provide a wealth of additional `free' information that can
be used to determine the timestep. The fourth order Hermite integrator, for
example, gives us $\dot{\vec{a}}$, $\ddot{\vec{a}}$ and $\dddot{\vec{a}}$ (the
latter two from finite differences). Such considerations have motivated
higher-order time-stepping criteria.

For example, an immediately obvious choice for a higher-order criterion is to
set the time step based on the truncation error in the Hermite expansion:
\begin{equation}
  \label{eqn:truncationstep}
  \Delta t_i = \big(\eta |\vec{a}_i|/|\ddot{\vec{a}}_i|\big)^{1/2}
  \qquad \text{or} \qquad
  \Delta t_i = \big(\eta |\vec{a}_i|/|\dddot{\vec{a}}_i|\big)^{1/3}
\end{equation}
or to use the error in the predictor step, as suggested by
\citet{NitadoriMakino2008}:
\begin{equation}
  \label{eqn:prederrstep}
  \Delta t_i = \Delta t_{\mathrm{old},i} \big(\eta
  |\vec{a}_i|/|\vec{a}_i - \vec{a}_{p,i}|\big)^{1/p}
\end{equation}
where $p$ is the order of the expansion, $\vec{a}_{p,i}$ the predicted
acceleration, and $\Delta t_{\mathrm{old},i}$ the previous timestep. However,
while such criteria seem sensible, they are all out-performed by the seemingly
mystic \citet{Aarseth2003} criterion
\begin{equation}
  \Delta t_i = \left(\eta \frac{|\vec{a}_i||\ddot{\vec{a}}_i| +
    |\dot{\vec{a}}_i|^2}{|\dot{\vec{a}}_i||\dddot{\vec{a}}_i| +
    |\ddot{\vec{a}}_i|^2}\right)^{1/2}
  \label{eqn:aarsethstep}
\end{equation} 
with $\eta \sim 0.02$ (which can be generalised for higher-order schemes, see
\citep{Makino1991,NitadoriMakino2008}).

The success of equation~(\ref{eqn:aarsethstep}) probably lies in the fact that
it conservatively shrinks the time step if either $\ddot{\vec{a}}$ or
$\dddot{\vec{a}}$ are large compared to the smaller derivatives, \emph{and} it
requires no knowledge of the previous
timestep. Equations~(\ref{eqn:truncationstep}) give poorer performance because
they do not use information about all known derivatives of
$\vec{a}$. Equation~(\ref{eqn:prederrstep}) gives poorer performance for large
timesteps. It too uses information about all calculated derivatives of
$\vec{a}$ (since it is based on the error between predicted and true
accelerations). But the problem is that it relies on the previous timestep for
its normalisation. If $\Delta t_{\mathrm{old},i}$ is too large, the criteria
will not respond fast enough, leading to overly large timesteps and large
energy losses \citep{NitadoriMakino2008}.

The above suggests that a conservative `truncation' error-like criteria that
encompasses all derivatives of $\vec{a}$ might perform at least as well as the
Aarseth criteria while being (perhaps) more theoretically satisfying:
\begin{equation}
  \label{eqn:justinstep}
  \Delta t_i = \mathrm{min}\left\{\left(\eta
    \frac{|\vec{a}_i|}{|\dot{\vec{a}}_i|}\right),\left(\eta^2
    \frac{|\vec{a}_i|}{|\ddot{\vec{a}}_i|}\right)^{1/2}, \left(\eta^3
    \frac{|\vec{a}_i|}{|\dddot{\vec{a}}_i|}\right)^{1/3}, ... , \left(\eta^p
    \frac{|\vec{a}_i|}{|{\vec{a}}^{(p)}_i|}\right)^{1/p} \right\}
\end{equation}
where $p$ is the highest order of $\vec{a}$ calculated by the integrator; and
$\vec{a}^{(p)}$ is the $p^\mathrm{th}$ derivative of $\vec{a}$.

In the right panel of Fig.~\ref{fig:force_schematic}, we compare 4th order
Hermite integrators with different variable timestep criteria for a Kepler
orbit problem with $e=0.9$. The black curve shows results for a fixed timestep
with $\Delta t=0.001$ periods; the red curve shows results using a
variable timestep and the Aarseth criteria (equation~\ref{eqn:aarsethstep});
and the red dotted curve shows results using
equation~(\ref{eqn:justinstep}). For the Aarseth criteria we use $\eta =
0.02$; for our new criteria in equation~(\ref{eqn:justinstep}) we set $\eta$
in all cases such that exactly the same number of steps are taken over ten
orbits as for the Aarseth criteria. The `truncation' error-like criteria
(equation \ref{eqn:justinstep}) appears to give very slightly improved
performance for the same cost. However, whether this remains true for full
$N$-body applications remains to be tested.

The above timestep criteria have been well tested for a wide range of problems
and so appear to work well---at least for the types of problem for which they
were proposed. However, there remains something unsatisfying about all of
them. For some, changing the velocity or potential can alter the timestep and,
with the possible exception of the criterion by \citet{ZempEtAl2007}, all are
affected by adding a constant to the acceleration. This is unsatisfactory,
since the internal dynamics of the system is not altered by any of these
changes. Applying a constant uniform acceleration, generated for example by an
external agent, to a star cluster is allowed by the Poisson equation and does
not alter the internal dynamics, and thus should not drastically alter the
timesteps. Only if the externally generated acceleration varies across the
cluster does it affect its internal dynamics, an effect known as tides.
This suggests 
\begin{equation}
  \label{eqn:walterstep}
  \Delta t_i = \left(\eta/||(\vec{\nabla}\!\vec{a})_i||\right)^{1/2}
\end{equation}
where $\vec{\nabla}\!\vec{a}$ is the gradient of the acceleration and
$||\cdot||$ denotes the matrix norm. Remarkably, for the Kepler problem this
agrees with equation~(\ref{eqn:leapfrogstep}), while for isolated systems with
power-law mass profiles it is very similar to
equation~(\ref{eqn:idealstep}). However, computing the gradient of $\vec{a}$
merely for the sake of the timestep seems extravagant.

\subsubsection{Close encounters and regularisation} 
A key problem when modelling collisional dynamics is dealing with the divergence
in the force for $\vec{x}_i\to\vec{x}_j$ in equation~(\ref{eqn:direct}),
requiring prohibitively small timesteps (or large errors) with any of the above
schemes. Consider our simple Kepler orbit problem. For a timestep criteria as in
equation~(\ref{eqn:leapfrogstep}), this gives a timestep at pericentre $r_p$ of
$\Delta t^2 \propto r_p^3/GM$. Thus, for increasingly eccentric orbits, the
timesteps will rapidly shrink, leading to a few highly eccentric particles
dominating the whole calculation. To avoid this problem, collisional $N$-body
codes introduce \emph{regularisation} for particles that move on tightly bound
orbits. The key idea is to use a coordinate transformation to remove the force
singularity, solve the transformed equations, and then transform back to
physical coordinates. Consider the equations of motion for a perturbed two-body
system with separation vector $\vec{R} = \vec{x}_1 - \vec{x}_2$ (using
$R\equiv|\vec{R}|$):
\begin{equation}
  \ddot{\vec{R}} = -G(m_1 + m_2)\frac{\vec{R}}{R^3} + \vec{F}_{12},
\end{equation} 
where $\vec{F}_{12} = \vec{F}_1-\vec{F}_2$ is the external perturbation. This,
of course, still has the singularity at $R=0$. Now, consider the time
transformation $\mathrm{d}t=R \mathrm{d}\tau$:
\begin{equation}
  \label{eqn:timetrans}
  \vec{R}'' = \frac{1}{R}R'\vec{R}' - G(m_1 + m_2)\frac{\vec{R}}{R} +
  R^{2}\vec{F}_{12}
\end{equation}
where $'$ denotes differentiation w.r.t.\ $\tau$. Note that we have removed the
$R^{-2}$ singularity in the force, but gained another in the term involving
$R'$. To eliminate that, we must also transform the coordinates. The current
transformation of choice is the \emph{Kustaanheimo--Stiefel} (K-S)
transformation \citep{KustaanheimoEtAl1965, Yoshida1982, Aarseth2003}, which
requires a move to four spatial dimensions. We introduce a dummy extra dimension
in $\vec{R}=(R_1,R_2,R_3,R_4)$, with $R_4=0$, and transform this to a new four
vector $\vec{u}=(u_1,u_2,u_3,u_4)$ such that $\vec{R} =
\mathcal{L}(\vec{u})\vec{u}$, with:
\begin{equation}
  \mathcal{L} = \left[
    \begin{array}{rrrr}
      u_1 & -u_2 & -u_3 & u_4 \\
      u_2 & u_1 & -u_4 & -u_3 \\
      u_3 & u_4 & u_1 & u_2 \\
      u_4 & -u_3 & u_2 & -u_1
    \end{array}
    \right]
\end{equation}
The inverse transformation is non-unique, since one of the components of
$\vec{u}$ is arbitrary. In general, we may write:
\begin{subequations}
  \label{eqn:uutrans}
  \begin{equation}
    \label{eqn:u1u2trans}
    u_1^2 = \frac{1}{2}(R_1+R)\cos^2\!\psi ; \qquad u_2 = \frac{R_2 u_1 + R_3
      u_4}{R_1+R}
  \end{equation}
  \begin{equation}
    \label{eqn:u4u3trans}
    u_4^2 = \frac{1}{2}(R_1+R)\sin^2\!\psi ; \qquad u_3 = \frac{R_3 u_1 - R_2
      u_4}{R_1+R}
  \end{equation}
\end{subequations}
where $\psi$ is a free parameter. It is a straightforward exercise to verify
that equations~(\ref{eqn:uutrans}) satisfy the transformation equation $\vec{R}
= \mathcal{L}(\vec{u})\vec{u}$. We also require a transformation between the
velocities $\dot{\vec{R}}$ and $\vec{u}'$. Writing $\vec{R}' =
\mathcal{L}(\vec{u}')\vec{u} + \mathcal{L}(\vec{u})\vec{u}' = 2
\mathcal{L}(\vec{u})\vec{u}'$, and using the relation
$\mathcal{L}^\mathrm{T}\mathcal{L} = R\vec{I}$ gives:
\begin{equation}
  \vec{u}' = \frac{1}{2} \mathcal{L}^\mathrm{T} \frac{\vec{R}'}{R} = \frac{1}{2}
  \mathcal{L}^\mathrm{T} \vec{\dot{R}}
\end{equation}
where the last relation follows from the time transformation $\mathrm{d}t = R
\mathrm{d}\tau$. Substituting the K-S coordinate transform into
equation~(\ref{eqn:timetrans}) gives \citep{Aarseth2003}:
\begin{subequations}
  \label{eqn:ks}
  \begin{equation}
    \vec{u}'' - \frac{1}{2}E\vec{u} = \frac{1}{2} R \mathcal{L}^{\mathrm{T}}
    \vec{F}_{12}
    \label{eqn:ksmotion}
  \end{equation}
  where $E$ is the specific binding energy of the binary, which is evolved as:
  \begin{equation}
    E' = 2 \vec{u}' \cdot \mathcal{L}^{\mathrm{T}} \vec{F}_{12}.
    \label{eqn:ksenergy}
  \end{equation}
\end{subequations}
(Note that the transformed time is given by $t' = |\vec{u}|^2$, which follows
from equation \ref{eqn:timetrans}.) We can now see two important
things. Firstly, there are no longer any coordinate singularities in
equations~(\ref{eqn:ks}). Secondly, in the absence of an external field
($\vec{F}_{12} = 0$), $E = \mathrm{const.}$ and our transformed equations
correspond to a simple harmonic oscillator.

We can evolve the above regularised equations of motion using the Hermite scheme
(\S\ref{sec:hermite}), so long as we can calculate $\vec{u}'''$ and $E''$. These
follow straightforwardly from the transformed time derivatives of
equations~(\ref{eqn:ks}):
  \begin{equation}
    \vec{u}''' = \frac{1}{2}\left(E' \vec{u} + E\vec{u}' + R' \vec{Q} +
    R\vec{Q}'\right),
    \qquad
    E'' = 2\vec{u}'' \cdot \vec{Q} + 2\vec{u}' \vec{Q}'
  \end{equation}
where $\vec{Q} = \mathcal{L}^\mathrm{T} \vec{F}_{12}$ describes the external
interaction term.

In Figure \ref{fig:force_schematic}(c), we show results for a Kepler orbit with
eccentricity $e=0.9$ integrated over 100 orbits using the K-S regularisation
technique (blue). We use a Hermite integrator with variable timesteps, and
timestep criterion~(\ref{eqn:justinstep}). For as many force calculations as the
variable timestep Hermite integration scheme, the results are over 100 times
more accurate. This is why K-S regularisation has become a key element in modern
collisional $N$-body codes.

K-S regularisation as presented above works only for a perturbed binary
interaction. However, it is readily generalised to higher order interactions
where for each additional star, we must transform away another potential
coordinate singularity \citep{AarsethZare1974,Heggie1974,Aarseth2003}. In
practice, this means introducing $N$ coupled K-S transformations, which requires
$4N(N-1)+1$ equations, making extension to large $N$ inefficient. For this
reason, \emph{chain regularisation} has become the state-of-the art
\citep{MikkolaAarseth1990,MikkolaAarseth1993}. The idea is to regularise only
the close interactions between the $N$ particles, rather than all inter-particle
distances, which reduces the number of equations to just $8(N-1)+1$, paving the
route to high $N$. For interactions involving large mass ratio, other
regularisation techniques can become competitive with the K-S chain
regularisation \citep{MikkolaAarseth2002}. This is particularly
important for interactions between stars and supermassive black holes.

\subsection{Recent numerical developments}
\label{sec:collisionalrecent} 
Many of the key algorithmic developments for collisional $N$-body simulations
were advanced very early on in the 1960's and 1970's \citep{Aarseth1963,
  AhmadCohen1973, AarsethZare1974, Heggie1974}. As a result, the field has been
largely driven by the extraordinary improvement in hardware. From the early
1990's onwards, the slowest part of the calculation -- the direct $N$-body
summation that scales as $N^2$ -- was moved to special hardware chips called
{\tt GRAPE} processors (GRAvity PipE; \citet{1990CoPhC..60..187I}). The latest
{\tt GRAPE-6} processor manages an impressive $\sim 1$\,Teraflop
\citep{2003PASJ...55.1163M}, allowing realistic simulations of star clusters
with up to $10^5$ particles \citep{2003MNRAS.340..227B}. However, to move
toward the million star mark (relevant for massive star clusters), several {\tt
  GRAPE} processors must be combined in parallel. This became possible only very
recently with the advent of the {\tt GRAPE-6A} chip
\citep{2005PASJ...57.1009F}. The {\tt GRAPE-6A} is lower performance (and
cheaper) than the {\tt GRAPE-6}, but specially designed to be used in a parallel
cluster. Such a cluster was recently used by \citet{HarfstEtAl2007} to model a
star cluster with $N=4\times10^6$.

The {\tt GRAPE} processors have been invaluable to the direct $N$-body
community, and with the recently developed {\tt GRAPE-DR}, they will continue
to drive the field for some time to come \citep{2008IAUS..246..457M}. However,
concurrent with the further development of the {\tt GRAPE} chips, significant
interest is now shifting towards Graphical Processor Units (GPUs) for hardware
acceleration. This is driven primarily by cost. Even the smaller and cheaper
{\tt GRAPE-6A} costs several thousand dollars at the time of writing and
delivers $\sim 150$ GigaFlops of processing power. By contrast GPUs deliver
$\sim 130$\,GigaFlops for just a couple of hundred dollars. The advent of a
dedicated $N$-body library for GPUs makes the switch to GPUs even easier
\citep{2009NewA...14..630G}. Whether the future of direct $N$-body
calculations lies in dedicated hardware, or GPUs remains to be seen. A third
way is entirely possible if new algorithms can make the force calculations
more efficient. We discuss the prospects for this, next.

\subsection{Critique and numerical alternatives}
\label{sec:collcritique}
One of the main problems with contemporary $N$-body codes for collisional
stellar systems is their great similarity. An immediate consequence is that the
usual method for the validation of simulation results (see \S\ref{sec:valid}) by
comparing independent approaches is hardly possible. In addition, splitting the
force computation into near and far field components is intimately connected
with the time integration such that one cannot simply change one without the
other.

The requirement for an accurate force computation by no means implies the need
for the costly brute-force approach currently employed.  Alternatively, an
approximative method with high accuracy, such as the fast-multipole method
(see \S\ref{sec:tree:fmm}), requires only $\mathcal{O}(N)$ instead of
$\mathcal{O}(N^2)$ operations to compute the forces for \emph{all} $N$
particles.

Time integration could perhaps also be improved. The Hermite scheme typically
employed is neither symplectic nor time-reversible. While this does not
necessarily imply that the time integration method causes uncontrolled
errors, it would certainly be desirable to compare to an alternative
method. One interesting option is to use fourth-order forward symplectic
integrators (\citep{ChinChen2005}, see also footnote~\ref{foot:4symp}) in
conjunction with a time-symmetric method for adapting the time steps,
resulting in an overall time-reversible scheme.

\subsection{Alternatives to N-body simulations}
\label{sec:collalternative}
Collisional stellar systems are typically in dynamical, or virial, equilibrium
where their overall properties, such as the mean density and velocity
distributions of stars, remain unchanged over dynamical time scales. However,
owing to stellar encounters, the systems evolves on much longer time
scales. $N$-body methods follow the dynamics on all time scales and thus do
not exploit the fact that the system is almost in dynamical equilibrium.

An alternative is to use a mean-field approach where a dynamical equilibrium
is evolved to another dynamical equilibrium, using some prescription for the
processes, such as stellar encounters, driving this evolution. This is the
gist of Fokker-Planck codes, which approximately solve the collisional
Boltzmann equation
\begin{equation}
  \label{eqn:colboltz} 
  \frac{\mathrm{d}f}{\mathrm{d}t}=
  \frac{\partial f}{\partial t}
  + \dot{\vec{x}}\cdot\frac{\partial f}{\partial\vec{x}}
  - \frac{\partial\Phi_{\mathrm{tot}}}{\partial\vec{x}}\cdot
  \frac{\partial f}{\partial\dot{\vec{x}}} = \Gamma[f].
\end{equation}
Here $f(\vec{x},\dot{\vec{x}},t)$ is the density of stars in six-dimensional
phase-space $\{\vec{x},\dot{\vec{x}}\}$, also known as the distribution
function (see \S\ref{sec:less:eqn} for more on this). The \emph{encounter
  operator} $\Gamma[f]$ describes the interaction between stars. In the limit
$\Gamma[f]\rightarrow 0$, equation~(\ref{eqn:colboltz}) recovers the
collisionless Boltzmann equation~(\ref{eq:cbe}). In general $\Gamma[f]$ is a
complicated non-trivial functional of $f(\vec{x},\dot{\vec{x}},t)$. When
replacing $\Gamma[f]$ with an approximation obtained using certain simplifying
assumptions, we obtain the Fokker-Planck equation (FPE, not given here, see
\citep{BinneyTremaine2008} for more details). The combined solution of the FPE
and the Poisson equation~(\ref{eq:Phi}) is still a formidable problem: solving
it in six phase-space coordinates plus time is simply unfeasible.  Several
methods have been employed to tackle this problem, very briefly summarised
below.

\vspace{-5mm}
\paragraph{Orbit averaging}
averages the net effect of the FPE over each orbit (assuming they are
regular), thus reducing from six to three dimensions. The resultant equations
may then be solved on a mesh. Following the pioneering work of
\citet{1979ApJ...234.1036C}, typically this has been done in two dimensions,
integrating the specific energy and $z$-component of the angular momentum
(e.g.~\citep{1995PASJ...47..561T}, we are unaware of any fully 3D calculation).

\vspace{-5mm}
\paragraph{Monte-Carlo}
methods, pioneered by \citet{1971Ap&SS..13..284H}, sample the stellar system
using tracer particles, whose trajectories are followed including not only the
background potential but also the diffusion in velocity according to the FPE.
Current implementations assume spherical symmetry to boost the resolution
(for a recent example see~\cite{2001A&A...375..711F}).

\vspace{-5mm}
\paragraph{Fluid Models}
take velocity moments of the FPE, resulting in fluid-like equations
\citep{1970MNRAS.147..323L}. This avoids the need to orbit-average the FPE,
but (1) implicitly assumes that scattering events are local and (2) requires
some assumption on the distribution function to close the hierarchy of moment
equations (i.e.\ an effective equation of state).

\medskip
Despite the large number of assumptions that go into workable Fokker-Planck
codes (independent local 2-body encounters only, assumed Coulomb logarithm,
etc.), the agreement with full $N$-body models is remarkable
\citep{1995MNRAS.272..772S, 1998HiA....11..591H,
  2008MNRAS.383....2K}. Currently, with such a large number of assumptions
Fokker-Planck codes mainly increase our understanding of the $N$-body
simulations, rather than act as an independent cross-check of the
results. However, given the rapid advance in computational power, it is perhaps
time to revisit this approach. One may also consider alternatives to the FPE
(but still based on equation~\ref{eqn:colboltz}), for example similar to the
\citet{Balescu1960}-\citet{Lenard1960} equation for plasma physics
\citep{Heyvaerts2010}. There remain significant numerical challenges to such
alternative approaches, but they hold the promise of a robust method for solving
collisional $N$-body dynamics that relies on very different assumptions to the
$N$-body method.

\subsection{Past, recent, and future astrophysical modelling}
\label{sec:coll:astr}
While the focus of this review rests firmly on numerical methods, we briefly
discuss our personal highlights of previous astrophysical results, as well as
recent and possible future developments.

Perhaps the earliest result was the numerical demonstration of
\emph{core-collapse} \citep{Aarseth1963}, which inspired the theory of
\emph{gravothermal-catastrophe} \citep{LyndenBellWood1968}. Since then the
increase in $N$ to nearly $10^5$ \citep{1994MNRAS.268..257G,
  1996MNRAS.282...19S, 1996ApJ...471..796M, BaumgardtEtAl2003} has confirmed
the onset of \emph{gravo-thermal oscillations} after core-collapse, first
discovered using a fluid code \citep{1983MNRAS.204P..19S}.  Exploring the
effect of binaries, several studies found that binaries may delay or even
reverse core collapse \citep{1990ApJ...362..522M, HeggieTrentiHut2006,
  2009PASJ...61..721T}.

Apart from being fascinating from a theorist's point of view, core collapse may
explain the rich variety of Globular Clusters observed in our Galaxy
\citep{1996AJ....112.1487H}, as well as potentially raising the central density
enough to promote stellar collisions. This latter process can seed the onset of
runaway growth, leading to the formation of intermediate mass black holes
(IMBHs)\footnote{There are a number of stellar-mass black-hole candidates
  \citep{1998ApJ...499..367B}, as well as super-massive black-hole candidates
  (${\sim}\,10^{6-9}$\,M$_\odot$, found at the centres of galaxies like our own
  Milky Way, e.g.\ \citep{1998AJ....115.2285M, 2010ApJ...721...26G}). However,
  there is no confirmed discovery of an IMBH with ${\sim}\,10^{3-4}$\,M$_\odot$,
  which are difficult to detect unambiguously, but would provide a missing link
  \citep{2001ApJ...551L..27M,2003ApJ...582..559V}.}  \citep{2002ApJ...576..899P,
  2004Natur.428..724P}. The firm detection IMBHs in star clusters in the basis
of dynamical evidence is quite challenging
\citep{GebhardtRichHo2002,2005ApJ...634.1093G} and requires understanding of the
collisional cluster dynamics \citep{2003ApJ...589L..25B,2005ApJ...620..238B}.

The most important force acting in collisional $N$-body systems is gravity, and
solving the gravitational force equation has been the focus of this review so
far. However, many other interesting physical processes are at play within the
stars. Stars evolve over time, moving off the main sequence onto the giant
branch and then eventually ending their lives as stellar remnants
\citep{Phillips1999}. Binary stars have even more complex lives, and all sorts
of interesting physics can occur due to mass transfer between them
\citep{2009MNRAS.395.1127C, 2011ApJ...726...66L}. In the very dense centres of
star clusters, physical stellar collisions can drive stellar evolution
\citep{1997ApJ...487..290S, 1998MNRAS.298...93B, BaumgardtKlessen2011}. And
finally, the dearth of observed gas in star clusters may provide interesting
constraints on stellar evolution models, in which case gas must also be included
in cluster models \citep{1977ApJ...211...77F, 2011MNRAS.411.1935P}.

The first attempt to mesh astrophysical processes with $N$-body models was
presented in 1987 by \citet{1987MNRAS.224..193T}. Since then, the number of
stars modelled and the complexity of the stellar and binary evolution models has
continued to grow \citep{2001MNRAS.321..199P, 2003ApJ...582L..21B,
  2005MNRAS.363..293H, 2009PASA...26...92C}. Recently, there has been a
dedicated drive to combine different physical simulation codes together within
one framework. This is the goal of the {\tt MODEST} collaboration and its
off-shoots\footnote{See also {\tt http://www.manybody.org/modest/}.}
\citep[Modelling and Observing DEnse STellar systems;][]{2003NewA....8..337H,
  2003NewA....8..605S, 2006NewA...12..201D, 2007jena.confE..63S,
  2010NewAR..54..163H}.

With ever-improving hardware and software, the challenges in modelling $N$-body
systems will shift from solving gravity, to building ever more realistic models
for the physics beyond the particle resolution (stellar evolution, binary
evolution, collisions, gas physics, and feedback processes from stars and
stellar remnants). Understanding these processes, and building believable
models, which can explain objects such as the 11 minute X-ray binary 4U\,1820-30
(see footnote~\ref{foot:4U1820-30}) as well as their distribution and evolution,
will become a new frontier of computational astrophysics in the coming decades.

\section{N-body methods for collisionless systems}
\label{sec:less}
In collisionless stellar systems the long-term effects of two-body encounters
are negligible. In other words, the gravitational potential governing the
stellar motions, which is actually the sum of many individual point-mass
potentials, is well approximated by a smooth mean potential
$\Phi(\vec{x},t)$. If unperturbed, a collisionless stellar system quickly
(within a few dynamical times) settles into dynamic equilibrium, when changes in
the mean density and potential become negligible and no further evolution
occurs. In this situation, the virial theorem
\begin{equation}
  \textstyle
  2T+W=0,\qquad\text{with}\qquad 
  T=\sum_i \frac{1}{2}\, m_i\,\dot{\vec{x}}_i^2
  \qquad\text{and}\qquad 
  W=\sum_i m_i \,\vec{x}_i{\,\cdot\,}\ddot{\vec{x}}_i
\end{equation}
holds and one speaks of \emph{virial equilibrium}. Because relaxation in the
thermodynamic sense does not occur, such virial equilibria have generically
non-Maxwellian and anisotropic velocity distributions, corresponding to a
tensor-like pressure in the fluid picture.

The main exponent of collisionless stellar systems are galaxies and systems of
galaxies (clusters and the universe as a whole), which have number densities too
small and dynamical times scales too long for stellar encounters to be
important. However, whereas for collisional systems external perturbations are
usually weak, such perturbations, including galaxy encounters and mergers, are
frequent and often significant for collisionless system. As a consequence,
collisionless systems are frequently perturbed away from equilibrium, resulting
in evolution to a new equilibrium. Another related process is secular evolution,
when the perturbation originates from an instability of the system itself.

\subsection{Equations governing collisionless stellar systems}
\label{sec:less:eqn}
Because the graininess of the stellar dynamics has negligible effect,
collisionless stellar systems are commonly described using continuum methods.
The fundamental quantity describing the state of the system at any time is its
\emph{distribution function} $f(\vec{x},\dot{\vec{x}},t)$, which is the mass
density of stars (in the continuum limit) in six-dimensional \emph{phase-space}
$\{\vec{x},\dot{\vec{x}}\}$ at time $t$. This is a significant simplification
compared to collisional systems (whose phase-space is $6N$ dimensional) and as a
consequence any correlations between particles, such as binaries and encounters,
are ignored\footnote{\label{foot:bbgky}In other words, the BBGKY hierarchy
  (after their discoverers Bogoliubov, Born, Green, Kirkwood, and Yvon, see also
  \citep{BinneyTremaine2008}) of the 1-body, 2-body, and higher order
  distribution functions is truncated at the lowest order.}. The continuous
spatial density and gravitational potential generated by the system are obtained
as
\begin{eqnarray}
  \label{eq:rho}
  \rho(\vec{x},t) &=& \int\mathrm{d}\dot{\vec{x}}\,f(\vec{x},\dot{\vec{x}},t),
  \\
  \label{eq:Phi}
  \Phi(\vec{x},t) &=& -G\int\mathrm{d}\vec{x}^\prime\,
  \frac{\rho(\vec{x}^\prime,t)}{|\vec{x}-\vec{x}^\prime|}
  = -G\int\!\!\!\!\int\mathrm{d}\dot{\vec{x}}\,\mathrm{d}\vec{x}^\prime\,
  \frac{f(\vec{x}^\prime,\dot{\vec{x}},t)}{|\vec{x}-\vec{x}^\prime|},
\end{eqnarray}
respectively. Because, according to Liouville's theorem, phase-space volume
remains constant along the flow and because of mass conservation, the ratio
$f=\mathrm{d} M/\mathrm{d}\{\vec{x},\dot{\vec{x}}\}$ is constant too, thus
satisfying
\begin{equation} \label{eq:cbe}
  0=\frac{\mathrm{d}f}{\mathrm{d}t}=
  \frac{\partial f}{\partial t}
  + \dot{\vec{x}}\cdot\frac{\partial f}{\partial\vec{x}}
  - \frac{\partial\Phi_{\mathrm{tot}}}{\partial\vec{x}}\cdot
  \frac{\partial f}{\partial\dot{\vec{x}}},
\end{equation}
known as the \emph{collisionless Boltzmann equation} (CBE). Here, we have used
the equation of motion $\ddot{\vec{x}}=-\vec{\nabla}\Phi_{\mathrm{tot}}$ with
the total potential $\Phi_{\mathrm{tot}}=\Phi+\Phi_{\mathrm{ext}}$, which
includes contributions from external agents not modelled by the distribution
function $f$. The evolution of a collisionless system is thus governed by the
CBE in conjunction with the Poisson equation~(\ref{eq:Phi}) and, possibly, an
external potential.

\begin{figure*}
  \centerline{
    \resizebox{120mm}{!}{\includegraphics{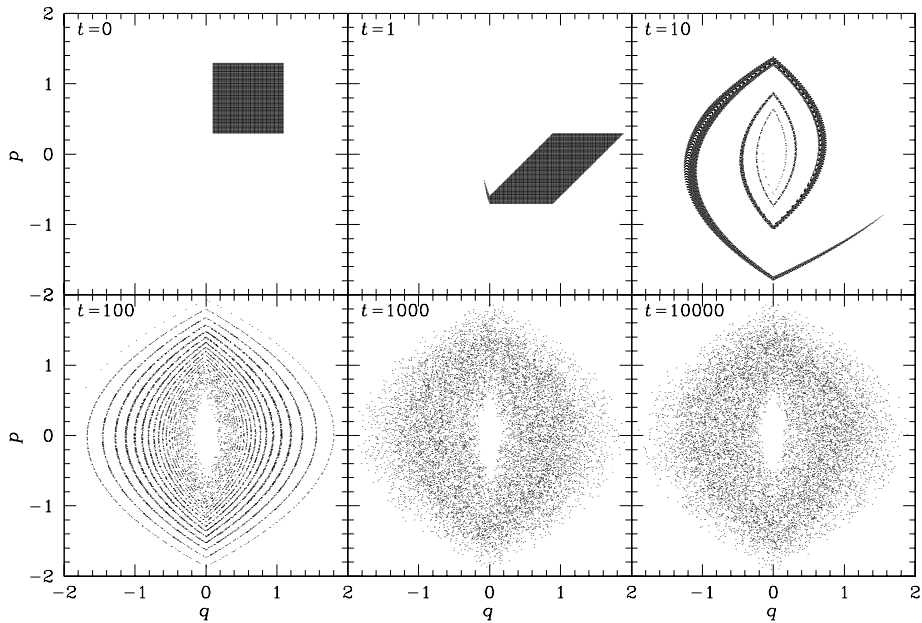}}
  }
  \caption{\small\label{fig:mixing} 
    The effect of mixing demonstrated with simple phase-mixing of $10^4$
    points in the Hamiltonian $H=\frac{1}{2}p^2+|q|$. This corresponds to
    massless tracer particles orbiting a central point mass in one-dimensional
    gravity (where phase-space is two-dimensional). The fine-grained
    distribution function is either 1 or 0, but at late times a smooth coarse
    grained distribution appears. Note that dynamical mixing in 3D
    self-gravitating systems is much faster and stronger than with this toy
    model.
  }
\end{figure*}
For equilibrium systems $\partial f/\partial t=0$ and thus
$\partial\Phi/\partial t=0$. In this case, any distribution function which
depends on the phase-space co-ordinates only through \emph{isolating integrals
  of motion}\footnote{Any function $I(\vec{x},\dot{\vec{x}})$, such that
  $\dot{I}=0$ and the constraint $I(\vec{x},\dot{\vec{x}})=I_0$ reduces the
  dimensionality of phase-space by one (isolates) is an isolating integral of
  motion. Examples are the orbital energy for static potentials, or the orbital
  angular momentum in case of a spherical potential.}  solves the CBE. This is
known as the \emph{Jeans theorem} and allows the construction of simple
equilibrium models. Unfortunately, the most general collisionless equilibria are
triaxial and even though most of their orbits are regular and respect three
isolating integrals, only the orbital energy allows simple treatment. Therefore,
the Jeans theorem is of practical usage only for systems of higher symmetry,
where the angular momentum (or one of its components) is also an isolating
integral.

For near-equilibrium systems, approximate solutions can be obtained by writing
$f=f_0+f_1$ (with $f_1\ll f_0$ and $f_0$ describing a given equilibrium model),
and ignoring the term of the CBE quadratic in $f_1$. The resulting linear
perturbation analysis gives insight into the stability properties of simple
equilibrium models. However, most galaxies frequently undergo strong
perturbations and considerable deviations from equilibrium, when these methods
fail and instead a full numerical treatment is required.

The direct numerical solution of the CBE, a non-linear PDE in seven dimensions,
is not feasible. This is not just because of the vastness of six-dimensional
phase-space and the inhomogeneity of $f$ (typically most of the mass resides in
a tiny fraction of the bound phase space). A much more severe problem is that
under the CBE $f$ develops ever stronger gradients, even when evolving towards
equilibrium. This is because $f$ is conserved along the flow, so that initial
fluctuations are not averaged away, but \emph{mixed} leading to ever thinner
layers of different phase-space density. These require ever higher numerical
resolution even though the system hardly evolves observably, as demonstrated in
Fig.~\ref{fig:mixing}. This figure also demonstrates that the mixing invalidates
the continuum limit: at late times the concept of the distribution function
becomes useless. An alternative is to \emph{average} over the fluctuations by
considering the `coarse-grained' phase-space density $\bar{f}$, a local mean of
$f$. However, for $\bar{f}$ no evolution equation exists\footnote{One option is
  to use the CBE also for $\bar{f}$ and combine it with some averaging or
  coarse-graining operation, depending on the level of fluctuations developing
  under the CBE. However, the choice of an appropriate coarse-graining operation
  is a fundamental problem with such an approach, partly because phase-space has
  no natural metric. One would have to use $\bar{f}$ itself to define the
  coarse-graining and it is not clear how this should be done, in particular for
  cold models, when $f$ is non-zero only on a hypersurface. This is the common
  situation with initial conditions for cosmological simulations.}.

\subsection{Numerics of collisionless N-body simulations}
\label{sec:less:num}
Fortunately enough, the above problems are easily overcome by the usage of
$N$-body techniques. The basic idea is to model the distribution function by an
ensemble of $N$ phase-space points $\{\vec{x}_i,\dot{\vec{x}}_i\}$, $i=1\dots N$
with weights $\mu_i$, which are randomly chosen to represent
$f(\vec{x},\dot{\vec{x}},t=0)$. The conservation of $f$ along the flow implied
by the CBE then means that the weights $\mu_i$ remain unchanged along each
trajectory, such that the task is reduced to integrating all $N$ trajectories.
Thus, in one sense an $N$-body code solves the CBE by the method of
characteristics for solving PDEs---the particle trajectories are the
characteristics of the CBE. In another sense, an $N$-body code is a Monte-Carlo
technique: any initial sample of $N$ phase-space points drawn from the same
distribution function at $t=0$ results in another, equally valid, $N$-body model
for the time evolution of $f(\vec{x},\dot{\vec{x}},t)$. An important consequence
is that the $N$ simulated particles are not modelling individual stars, nor is
it helpful to consider them as `super stars' of enormous mass. The only correct
interpretation is that the ensemble of all $N$ particles together represents the
continuous distribution function $f$ (and in fact provides an implied
coarse-graining dependant on the numerical resolution). Thus unlike the
situation for collisional $N$-body methods, the number $N$ of particles is a
numerical parameter, controlling the resolution and by implication the accuracy
for certain predictions of the model.

With the $N$-body method one has no knowledge of the distribution function $f$
at $t>0$, except at the phase-space positions of the $N$ particles (where $f$
remains at its original value by virtue of the CBE). However, this is not a
problem at all, not least because the structure of $f$ is not necessarily very
useful (see above), but also because any moments of the distribution function
can be \emph{estimated} from the $N$ particles in the usual Monte-Carlo way
\begin{equation} \label{eq:est:mom}
  \langle g \rangle(t) \equiv
  \int\!\!\!\!\int\mathrm{d}\vec{x}\,\mathrm{d}\dot{\vec{x}}\,
  g(\vec{x},\dot{\vec{x}})\,f(\vec{x},\dot{\vec{x}},t)
  \approx \sum_i \mu_i\,g\big(\vec{x}_i(t),\dot{\vec{x}}_i(t)\big).
\end{equation}
Here, the function $g$ specifies the moment, for example $g=1$ obtains the
total mass, while $g=\frac{1}{2}\dot{\vec{x}}^2$ results in the total kinetic
energy. Such moments are in fact all one ever wants to know: all observable
properties, including the coarse-grained distribution function, are just
moments of $f$. However, one must not forget that this estimation procedure is
always subject to shot noise, the amplitude of which depends on the number of
particles effectively contributing, and hence on the numerical resolution and
the width of the moment function $g$.

\subsection{Cosmological N-body simulations} 
\label{sec:cosmo}
For cosmological simulations, in principle we should switch to general
relativistic (GR) equations of motion. However, we can take advantage of a very
useful approximation. On large scales, the Universe is very nearly isotropic and
described well by a smooth Friedmann-Lama\^itre-Robertson-Walker (FLRW)
spacetime \citep{2011ApJS..192...18K}. However, on small scales the Universe
must tend towards a locally inertial frame in which Newton's laws are
valid. This fact can be used to show that the cosmological expansion has
essentially no effect on the local dynamics, even on galaxy cluster scales
\citep{1998ApJ...503...61C}. Thus, we may think of the Universe as being filled
with a self-gravitating fluid that locally obeys Newton's law of gravitation,
while expanding as an FLRW metric on large scales\footnote{Recently, doubts have
  been raised about the validity of this approximation. Various authors
  suggested that local inhomogeneities affect the large-scale dynamics, for
  example mimicking cosmic acceleration which traditionally has been attributed
  to dark energy \citep{1997PhRvL..78.1624M, 2000PhRvD..62d3525B}. This seems to
  be unlikely \citep{2006CQGra..23..235I}, though the necessary corrections due
  to inhomogeneities could affect attempts to use $N$-body simulations to
  precisely determine cosmological parameters
  \citep{2011arXiv1104.0730E}.}. This leads to the following Hamiltonian for any
orbiting test-particle \citep{Peebles1980, 2005MNRAS.364.1105S}:
\begin{equation} \label{eq:Ham:cosmo}
  H = \frac{\vec{p}_i^2}{2 m_i a^2} + \frac{m_i}{2a}\,\Phi(\vec{x}_i),
\end{equation} 
where $\vec{x}_i$ and $\vec{p}_i = a^2 m_i \dot{\vec{x}}_i$ are now
\emph{co-moving} canonical coordinates (with respect to the isotropic, smooth,
expanding background spacetime); $a\equiv a(t)$ is the \emph{scalefactor} that
follows from the FLRW model; and the equations of motion follow in the usual way
from Hamilton's equations. Since we are now working in co-moving space, the
potential is really the \emph{peculiar} potential with respect to the smooth
background; while the forces are peculiar forces with respect to the
expansion. Note that the above Hamiltonian now explicitly depends on time and
thus does not conserve energy. This is the standard problem of ambiguous energy
conservation in GR \citep[for a discussion see e.g.][]{Peebles1980}.

The usual method for approximating isotropy and homogeneity of the universe on
large scales is to simulate a small cubic patch of size $L$ of the universe (to
satisfy the `local-Newtonian' approximation), and apply periodic boundary
conditions in co-moving co-ordinates
\begin{equation}
  \label{eq:Phi:cosmo}
  \Phi(\vec{x},t) = -G \sum_{\muin{n}} \int\mathrm{d}\vec{x}^\prime\,
  \frac{\rho(\vec{x}^\prime+\muin{n}L,t)}{|\vec{x}-\vec{x}^\prime-\muin{n}L|}
  = -G \sum_{\muin{n}} 
  \int\!\!\!\!\int\mathrm{d}\dot{\vec{x}}\,\mathrm{d}\vec{x}^\prime\,
  \frac{f(\vec{x}^\prime+\muin{n}L,\dot{\vec{x}},t)}
       {|\vec{x}-\vec{x}^\prime-\muin{n}L|},
\end{equation}
where the sum over $\muin{n}=(n_x,n_y,n_z)$ accounts for all periodic replica.
In practice, the periodic sum is approximated using \citet{Ewald1921}'s
method, which was originally invented for solid-state physics and imported to
this field by \citet{HernquistBouchetSuto1991} (but note an error in their
eq.~2.14b as pointed out by \citep{Klessen1997}). Alternatively, Fourier
methods, which naturally provide periodic boundary conditions, can be used,
see \S\ref{sec:force:grid}.

\subsection{Force softening} \label{sec:soft}
In order to obtain the particle trajectories, one just has to integrate the
equations of motion $\ddot{\vec{x}}_i = -\vec{\nabla} \Phi_{\mathrm{tot}}
(\vec{x}_i)$ for all $N$ particles. This is usually done using the leapfrog
integrator possibly with individual timesteps arranged in the block-step scheme
(\S\ref{sec:leapfrog}). The self-potential $\Phi$ must be estimated from the
positions and masses of the particles themselves. By virtue of
equations~(\ref{eq:Phi}) or (\ref{eq:Phi:cosmo}), $\Phi$ is a moment of $f$ and
so we can estimate it. However, the straightforward application of
equation~(\ref{eq:est:mom}) results simply in the equations of
motion~(\ref{eqn:direct}) for a collisional system with $N$ stars of masses
$\mu_i$ (and possibly immersed in an external gravitational field). This is not
what we want to model and, moreover, integrating these equations numerically is
rather difficult, as we have discussed in \S\ref{sec:timestep} above.

The problem is that the function $g=-G/|\vec{x}-\vec{x}^\prime|$ for estimating
the potential via equation~(\ref{eq:est:mom}) is quite localised and even
diverges for $\vec{x}\to\vec{x}^\prime$, such that shot noise becomes a serious
issue with the simple Monte-Carlo integration of
equation~(\ref{eq:est:mom}). This is closely related to estimating the spatial
density $\rho$. The application of equation~(\ref{eq:est:mom}) with $g=
\delta(\vec{x} -\vec{x}^\prime)$ obtains a sum of $\delta$-functions at the
instantaneous particle positions, consistent with the
accelerations~(\ref{eqn:direct}) from the Poisson equation $4\pi
G\rho=-\vec{\nabla}{\,\cdot\,}\vec{a}$. The problem of estimating a smooth
density from scattered data points (in several dimensions) is generic to many
applications in science. A number of solutions are known. One of them is to
widen the $\delta$-spikes in the density estimate to a finite size, resulting in
the estimator \citep{Silverman1986}
\begin{equation} \label{eq:est:rho}
  \hat\rho(x) = \sum_i \frac{\mu_i}{\epsilon^3}\,
  \eta\!\left(\frac{|\vec{x}-\vec{x}_i|}{\epsilon}\right),
\end{equation}
where $\eta(\xi)$ is the dimensionless kernel function (normalised to unit
integral) and $\epsilon$ the \emph{softening} length. The corresponding
estimator for the potential then follows by application of the Poisson equation,
or equivalently, by replacing the Greens function $-G/|\vec{x}-\vec{x}^\prime|$
with the potential generated by a mass distribution of the density kernel
\begin{equation} \label{eq:est:phi}
  \hat\Phi(x) = - \sum_i \frac{G\mu_i}{\epsilon}\,
  \varphi\!\left(\frac{|\vec{x}-\vec{x}_i|}{\epsilon}\right)
\end{equation}
where $4\pi\eta=-\xi^{-2}(\xi^2\varphi')'$ (a prime denoting
differentiation). For example, the commonly used Plummer softening
\begin{equation} \label{eq:plummer}
  \hat\Phi(\vec{x}) = -G \sum_i
  \frac{\mu_i}{\sqrt{|\vec{x}-\vec{x}_i|^2+\epsilon^2}}
\end{equation}
corresponds to replacing each particle by a \citet{Plummer1911} sphere of
scale radius $\epsilon$ and mass $\mu_i$, i.e.\
\begin{equation}
  \eta(\xi) = (3/4\pi)\,(\xi^2+1)^{-5/2}
  \qquad\text{and}\qquad
  \varphi(\xi)=(\xi^2+1)^{-1/2}.
\end{equation}

\subsubsection{Softening as a method to suppress close encounters}
 \label{sec:soft:close}
The softening of gravity has two important aspects. First, at close distances
$r=|\vec{x}_i-\vec{x}|$ the force no longer diverges as $r^{-2}$, but for
$r\lesssim\epsilon$ actually decays to zero.  As a consequence, the force
estimated from all particles is a smooth and continuous function, such that
integrating the equations of motion is much simpler than for the un-softened
case (required with collisional $N$-body methods). This also means that the
effect and importance of close particle encounters is suppressed. This is
exactly what we want, because close encounters are not described by the one-body
distribution function $f$, but by the 2-body distribution function, the next in
the BBGKY hierarchy (see also footnote~\ref{foot:bbgky}). The minimum softening
length required to prevent large-angle deflections during close encounters is
\begin{equation} \label{eq:eps:2b}
  \epsilon_{2\mathrm{body}} \sim G\mu/\sigma^2
\end{equation}
with $\mu$ the particle mass and $\sigma$ the typical velocity dispersion
\citep{White1979}.

In other words, close encounters between the integrated trajectories are
numerical artifacts and by softening the forces we eliminate their (artificial)
effects and, at the same time, considerably simplify the task of time
integration compared to collisional $N$-body codes. Unfortunately, two-body
relaxation is driven not only by close encounters, but all octaves of impact
parameter contribute equally. Therefore, force softening does not, contrary to
some common opinion, much reduce the \emph{artificial} two-body relaxation
(only by $\sim2$, \citep{Theis1998}). This implies that artificial relaxation
effects may affect the high-density regions of $N$-body simulations (where
$t_{\mathrm{dyn}}$ and hence $t_{\mathrm{relax}}$ is shortest, see
equation~\ref{eq:t:relax}). Thus while we aim to model collisionless systems,
the simulations themselves may still suffer from small-angle deflections.

\subsubsection{Softening as optimal force estimation:
  the choice of the softening kernel} \label{sec:soft:est}
The second aspect of force softening is a systematic reduction of gravity at
close distances: in the limit of $N\to\infty$ but fixed $\epsilon$, the gravity
estimated by (\ref{eq:est:phi}) disagrees with that of the system modelled.
This \emph{bias} of the average $N$-body force is the price for the reduction in
shot noise (i.e.\ suppression of close encounters). The presence of this force
bias implies that simulation results on scales smaller than a few $\epsilon$ are
unreliable. 

The overall force error is a combination of this bias and the noise, measured by
the variance of the force estimate, and depends on the system modelled, the
number $N$ of particles, the softening length $\epsilon$ and the softening
kernel $\eta$. For small $\epsilon$, the bias and variance for the
\emph{estimated} force $\hat{\vec{F}}$ can be approximated analytically using a
local Taylor expansion \citep{Dehnen2001}
\begin{eqnarray} \label{eq:F:bias}
  \mathrm{bias}\{\hat{\vec{F}}(\vec{x})\} = 
  a_0\epsilon^2G\,\vec{\nabla}\rho(\vec{x}) +
  a_2\epsilon^4G\,\vec{\nabla}\vec{\nabla}^2\rho(\vec{x}) +
  \mathcal{O}(\epsilon^6)
  \qquad&\text{with}&\qquad
  a_k = \frac{(4\pi)^2}{(k+3)!}\int_0^\infty\mathrm{d}\xi\,\xi^{k+4}\eta(\xi),
  \\[2ex]
  \label{eq:F:var}
  N\mathrm{var}\{\hat{\vec{F}}(\vec{x})\} =
  b\,G^2M\,\epsilon^{-1}\rho(\vec{x}) +
  \mathcal{O}(\epsilon^0)\qquad\qquad\qquad\quad
  \qquad&\text{with}&\qquad
  b = (4\pi)^2 \int_0^\infty\mathrm{d}\xi\,\xi^2\,\eta(\xi)\,\varphi(\xi).
\end{eqnarray}
with $\rho(\vec{x})$ and $M$ the (smooth) density of the system modelled and its
total mass, respectively. Thus, the total force error,
$\mathrm{bias}\{\hat{\vec{F}}(\vec{x})\}^2+
\mathrm{var}\{\hat{\vec{F}}(\vec{x})\}$, becomes minimal for
$\epsilon_{\mathrm{opt}}\propto N^{-1/5}$, depending on the system and the
softening kernel. In particular, it is required that $\xi^5\eta(\xi)\to0$ as
$\xi\to\infty$ for the constant $a_0$ to be finite. For kernels, such as Plummer
softening (for which $\eta\propto \xi^{-5}$ at $\xi\gg1$), which do not
satisfy this condition, the expansion above does not work and the force bias
grows faster than quadratic.

While these static considerations may not be directly relevant for the dynamical
evolution of the $N$-body system, it appears best to reduce the force bias at
least to $\mathcal{O}(\epsilon^2)$ by using softening kernels with either finite
density support (corresponding to exact $1/r$ gravity and $\eta=0$ for
$r>\epsilon$) or a decay steeper than $\eta\propto \xi^{-5}$ at $\xi\to\infty$.

In view of equation (\ref{eq:F:bias}), one may even reduce the bias further by
designing softening kernels with $a_0=0$ \citep{Dehnen2001}. Such kernels have
$\eta<0$ over some radial range (usually at large $\xi$), and compensate the
rare under-estimation of gravity at small radii with a slight over-estimation at
larger radii. The resulting reduction in bias allows a larger softening length
to suppress the noise. However, equation~(\ref{eq:F:bias}) is applicable only
when the density $\rho(\vec{x})$ is smooth on scales smaller than
$\epsilon$. Since many collisionless systems have quite steep central density
gradients, where $\rho$ may even formally diverge like a power law, techniques
(like this) which require larger softening length appear less advisable.

\subsubsection{The choice of the softening length}
While $\epsilon_{2\mathrm{body}}$ (equation~\ref{eq:eps:2b}) represents a
minimum softening length required to suppress artificial large-angle
deflections, larger values simplify the time integration further and hence
reduce the computational costs. Another useful criterion is that the maximum
inter-particle force shall not exceed the typical mean-field strength. For the
situation of a single stellar system of mass $M$ and radius $R$, this translates
to $G\mu/\epsilon^2\lesssim GM/R^2$. Replacing $M=N\mu$, we find
\citep{PowerEtal2003}
\begin{equation}
  \epsilon_{\min} \sim R / \sqrt{N}.
\end{equation}
This is significantly larger than $\epsilon_{2\mathrm{body}}\sim2R/N$ (which
follows from equation~\ref{eq:eps:2b} with $\sigma^2\sim GM/2R$).  Most
practitioners are guided by these simple considerations, and practical
convergence studies regarding the best choice for $\epsilon$ are lacking. A
notable exception is the work of \citet{PowerEtal2003}, who suggest for
cosmological simulations $\epsilon\sim 4\epsilon_{\min}$.

\subsubsection{Adaptive individual softening}
\label{sec:adapt:eps}
The local resolution of an $N$-body system is determined by the particle number
density and an obvious idea is to use smaller softening lengths in high-density
regions. This leads to the concept of individual $\epsilon_i$ with
$\epsilon^2_{i\!j}=(\epsilon^2_i + \epsilon^2_j)/2$ the softening used in the
interaction between particles $i$ and $j$ (other symmetrisations are possible,
but this particular one allows an efficient approximation when using the tree
code \citep{SaitohMakino2010}). The individual softening lengths are adapted,
for example such that $\epsilon_i^3\hat{\rho}_i$ remains constant, equivalent to
the adaption of individual smoothing lengths in smoothed-particle hydrodynamics
(SPH, \citep{Price2010} see also Lodato \& Cossins, this issue).

When implementing such a scheme, two things must be ensured in order to
guarantee the validity of the $N$-body method. First, the adaption of the
softening lengths must be time-reversible. This can be achieved to sufficient
accuracy by a technique equivalent to that used in SPH \citep[appendix
  A1]{CullenDehnen2010}. Second, since $\epsilon_i$ depends (implicitly) on the
positions of particle $i$ \emph{and} its neighbours, the $N$-body force
$\partial\hat{\Phi}/\partial\vec{x}_i$ contains additional terms, which must be
included, as outlined by \citet{PriceMonaghan2007}, to preserve the Hamiltonian
character of the method and hence energy conservation.

\subsection{Force computation}
\label{sec:less:force}
In collisionless $N$-body methods the force is only ever an estimate, which
unavoidably carries with it an \emph{estimation error} (which is reduced by
force softening, see above). Therefore, we may as well use less accurate methods
for the actual calculation of the estimated forces than the computationally
expensive direct summation, i.e.\ the straightforward implementation of
equation~(\ref{eq:est:phi}) or (\ref{eq:plummer}). Because the computational
effort of any $N$-body method is always dominated by the calculation of the
gravitational forces, considerable effort has been invested into the design of
fast force calculation algorithms, resulting in many different methods, which we
describe in some detail below. All of these methods are substantially faster
than direct summation and together with the simplifications for the
time-integration due to force softening, allow $N$ to be $\sim4$ orders of
magnitude higher in collisionless than collisional $N$-body simulations.

\subsubsection{Approximating direct summation}
\label{sec:tree:fmm}
A number of methods are based on approximating the direct summation
\begin{equation} \label{eq:est:pot}
  \hat\Phi(\vec{x}_b) = - \sum_a \mu_a\;\phi(\vec{x}_b-\vec{x}_a)
\end{equation}
(corresponding to equation~\ref{eq:est:phi} with $G$ and $\epsilon$ absorbed
into $\phi$) by replacing the contributions from all particles within a local
group by a single expression.

\vspace{-4mm}
\paragraph{The tree code} was pioneered by \citet{BarnesHut1986} in 1986 and
uses a hierarchical spatial tree to define localised groups of
particles. Because stellar systems are often highly inhomogeneous, this is much
better than defining groups by cells of an equidistant mesh (when few cells
would contain most of the particles). With the usual oct-tree each cubic cell
containing fewer than $n_{\max}$ particles is split into up to eight child cells
of half their parent's size. This results in a tree-like hierarchy of cubic
nodes with the root box, containing all particles, at its bottom. The particles
within each of the tree nodes constitute a well-defined and localised group.
The approximation used in the tree code is formally obtained by Taylor expanding
the kernel function $\phi$ in $\vec{x}_a$ around some expansion centre
$\vec{z}_{\!A}$ of the group\footnote{Using the multi-index notation
  $\muin{n}\equiv(\mathsf{n}_x,\mathsf{n}_y,\mathsf{n}_z)$ with
  $\mathsf{n}_i\ge0$, $\mathsf{n}\equiv|\muin{n}|\equiv
  \mathsf{n}_x+\mathsf{n}_y+\mathsf{n}_z$,
  $\vec{r}^{\muin{n}}\equiv r_x^{\mathsf{n}_x} r_y^{\mathsf{n}_y}
  r_z^{\mathsf{n}_z}$, and $\muin{n}!\equiv
  \mathsf{n}_x\!!\,\mathsf{n}_y\!!\,\mathsf{n}_z\!!$.}
\begin{equation} \label{eq:phi:expand:a}
  \phi(\vec{x}_b-\vec{x}_a) \approx \sum_{|\muin{n}|\le p}\frac{1}{\muin{n}!}
  (\vec{x}_a-\vec{z}_{\!A})^{\muin{n}}\,\vec{\nabla}^{\muin{n}}
  \phi(\vec{x}_b-\vec{z}_{\!A}),
\end{equation}
where $p$ is the expansion order. By inserting this expansion
into~(\ref{eq:est:pot}) we obtain for the potential from the group $A$ the
\emph{multipole expansion}
\begin{equation} \label{eq:pot:tree}
  \hat\Phi_{\!A}(\vec{x}_b) = - \sum_{a\in A} \mu_a\;\phi(\vec{x}_b-\vec{x}_a)
  \approx - \sum_{|\muin{n}|\le p} M_{\muin{n}}(\vec{z}_{\!A})
  \,D_{\muin{n}}(\vec{x}_b-\vec{z}_{\!A})
\end{equation}
with the derivatives
$D_{\muin{n}}(\vec{r})\equiv\vec{\nabla}^{\muin{n}}\phi(\vec{r})$ and the
multipoles of group $A$ w.r.t.\ its expansion centre $\vec{z}_{\!A}$
\begin{equation} \label{eq:mm}
  M_{\muin{n}}(\vec{z}_{\!A}) = \sum_{a\in A} \mu_a
  \frac{(-1)^{\mathsf{n}}}{\muin{n}!}  (\vec{x}_a-\vec{z}_{\!A})^{\muin{n}}.
\end{equation}
For the un-softened case ($\phi=1/|\vec{r}|$), this series converges in the
limit $p\to\infty$ if
$|\vec{x}_b-\vec{z}_{\!A}|>\max_{a}\{|\vec{x}_a-\vec{z}_{\!A}|\}$, i.e.\ if
$\vec{x}_b$ is outside a sphere centred on $\vec{z}_{\!A}$ and containing all
$\vec{x}_a$.  Furthermore, if the centre of mass of the group $A$ is chosen as
its expansion centre $\vec{z}_{\!A}$, the dipole vanishes such that the
zero-order expansion ($p=0$) is actually first-order accurate. Because of its
great simplicity, this approach is in fact a common choice for practical
implementations of the tree code.

In a preparatory step, the multipole moments (\ref{eq:mm}) and sizes $w_{\!A}$
satisfying $w_{\!A}\ge\max_{a}\{|\vec{x}_a-\vec{z}_{\!A}|\}$ are computed for
each tree cell. (Both is best done recursively, exploiting the results from the
daughter cells via the shifting formula
\begin{equation} \label{eq:mm:shift}
  M_{\muin{n}}(\vec{z}+\vec{x})=\sum_{|\muin{k}|\le|\muin{n}|}
  \frac{\vec{x}^{\muin{k}}}{\muin{k}!}\,M_{\muin{n}-\muin{k}}(\vec{z}),
\end{equation}
sometimes called the `upward pass'.) The gravity at any position $\vec{x}$
from all the particles within tree cell $A$ is then simply approximated by
applying equation~(\ref{eq:pot:tree}) if\footnote{Ensuring convergence of the
  series. Curiously, some early implementations used for $w_{\!A}$ simply the
  linear size of the cubic cell, when
  $w_{\!A}\ge\max_{a}\{|\vec{x}_a-\vec{z}_{\!A}|\}$ is \emph{not} guaranteed
  and the approximated forces can be catastrophically wrong, resulting in the
  infamous `exploding galaxies' bug \citep{SalmonWarren1994}.}
$\theta w_{\!A}<r=|\vec{x}-\vec{z}_{\!A}|$ with some \emph{opening angle}
$\theta\le1$. Otherwise, the sum of the potentials obtained by applying the
same algorithm to the daughter cells is used (or, if the cell is a tree leaf,
from direct summation over all particles within the cell). Of course, the
force is computed as the derivative of $\hat\Phi$.

\begin{figure*}
  \centerline{
    \resizebox{58mm}{!}{\includegraphics{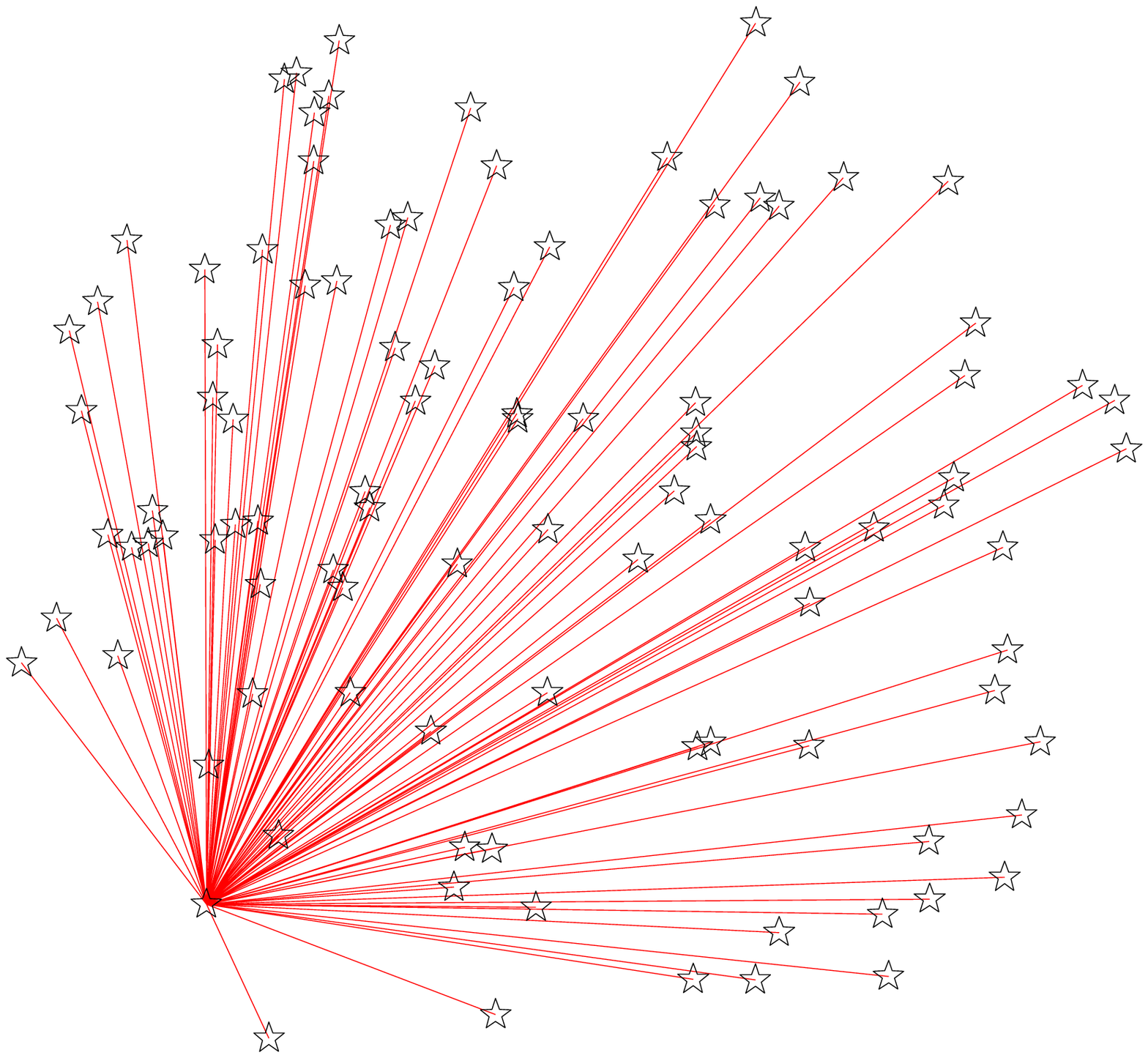}}
    \resizebox{59mm}{!}{\includegraphics{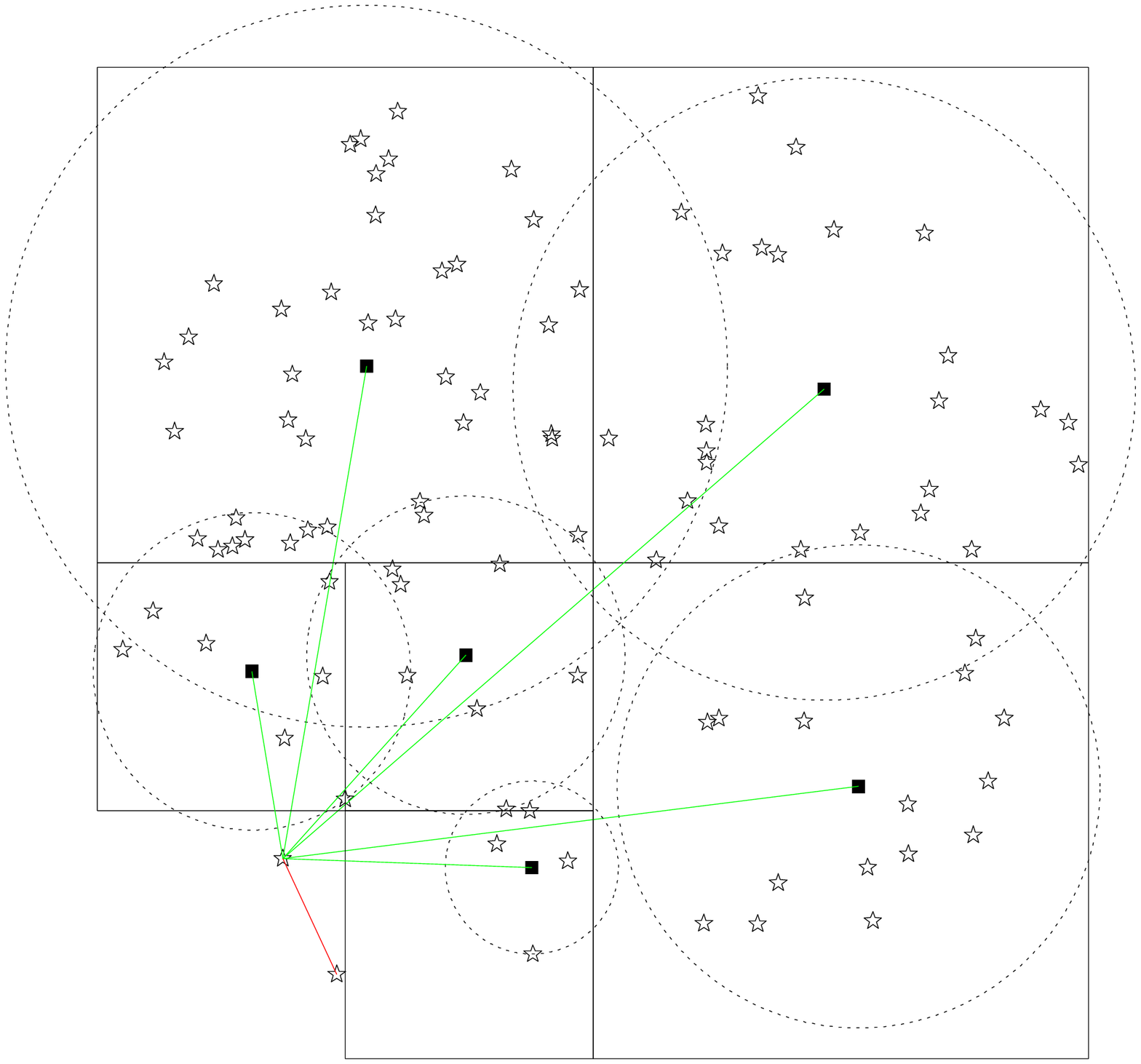}}
    \resizebox{59mm}{!}{\includegraphics{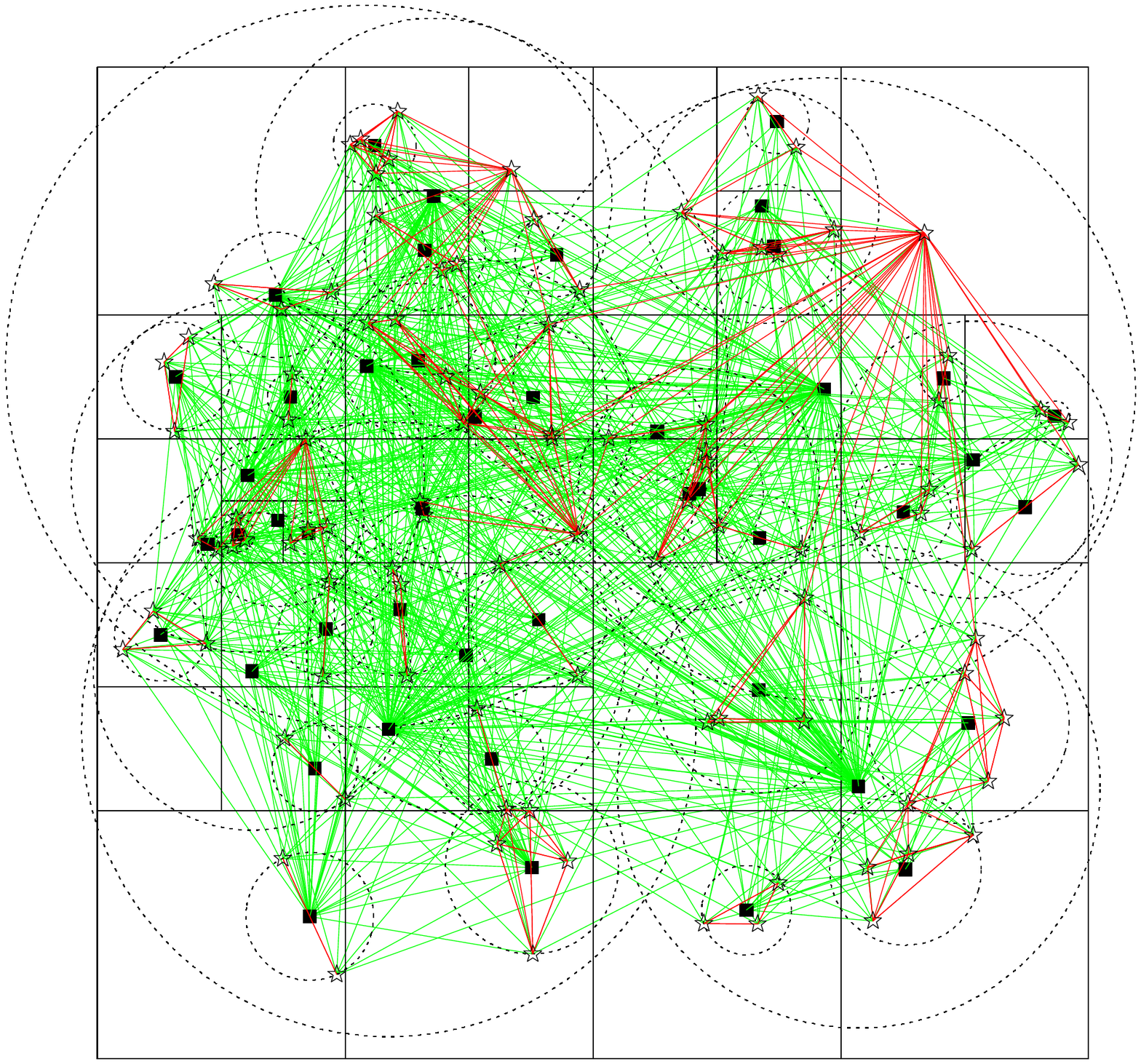}}
  }
  \caption{\small\label{fig:treecode} 
    \textbf{Left:} computation of the force for one of 100 particles (asterisks)
    in two dimensions (for graphical simplicity) using direct summation: every
    line corresponds to a single particle-particle force calculation.
    \textbf{Middle:} approximate calculation of the force for the same particle
    using the tree code. Cells opened are shown as black squares with their
    centres $\vec{z}$ indicated by solid squares and their sizes $w$ by dotted
    circles. Every green line corresponds to a cell-particle interaction.
    \textbf{Right:} approximate calculation of the force for all 100 particles
    using the tree code, requiring 902 cell-particle and 306 particle-particle
    interactions ($\theta=1$ and $n_{\max}=1$), instead of 4950
    particle-particle interactions with direct summation.
  }
\end{figure*}
Fig.~\ref{fig:treecode} demonstrates the working of the tree code and compares
it graphically to the direct summation approach, by applying both to the task of
computing the force for one of 100 particles (\emph{left} and \emph{middle}
panels). Obviously, the tree code requires much fewer calculations---it is
straightforward to show that the number of force computations per particle
scales like the depth of the tree, i.e.\ $\ln N$, such that the cost of
computing all $N$ forces is $\mathcal{O}(N\ln N)$.

\begin{figure}
  \centerline{
    \parbox[b]{80mm}{
      \resizebox{80mm}{!}{\includegraphics{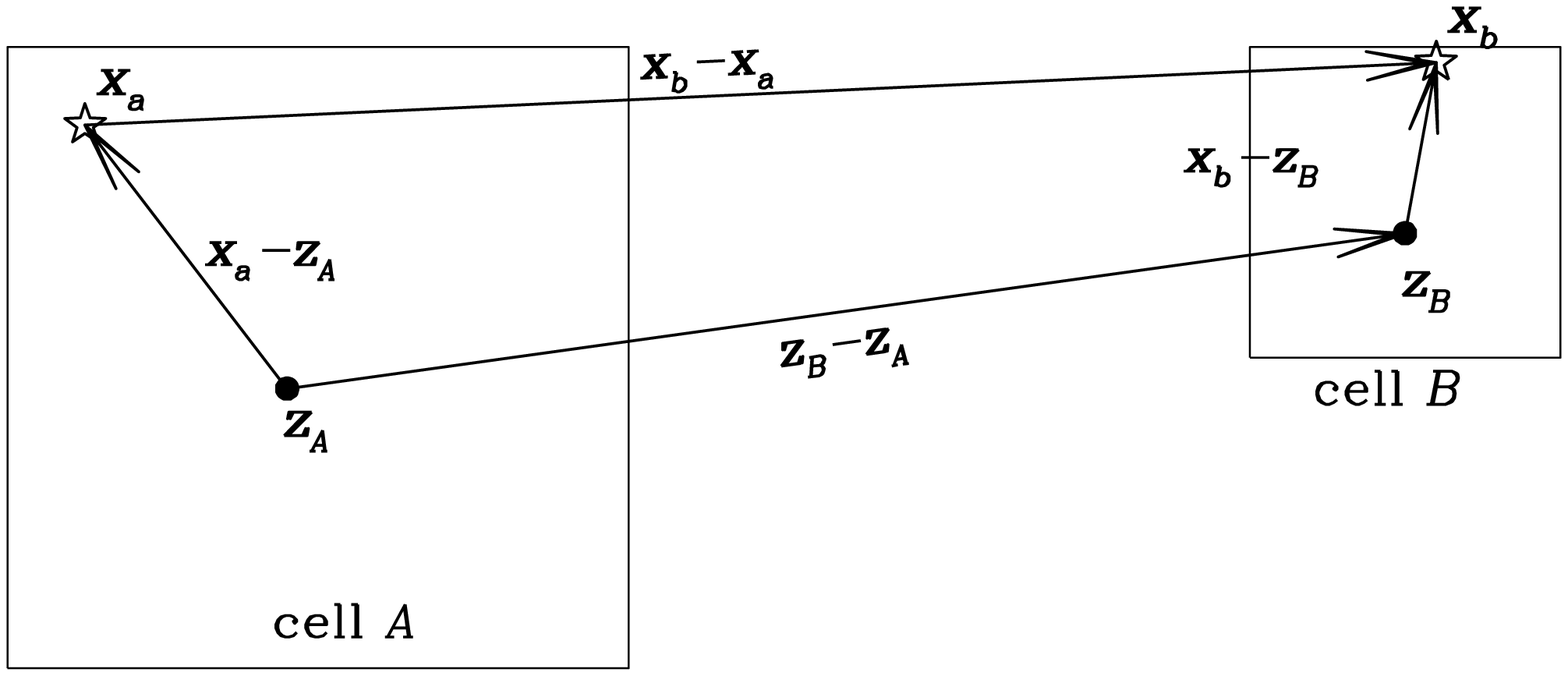}}
      \vspace{8mm}
    }
    \hspace{15mm}
    \resizebox{59mm}{!}{\includegraphics{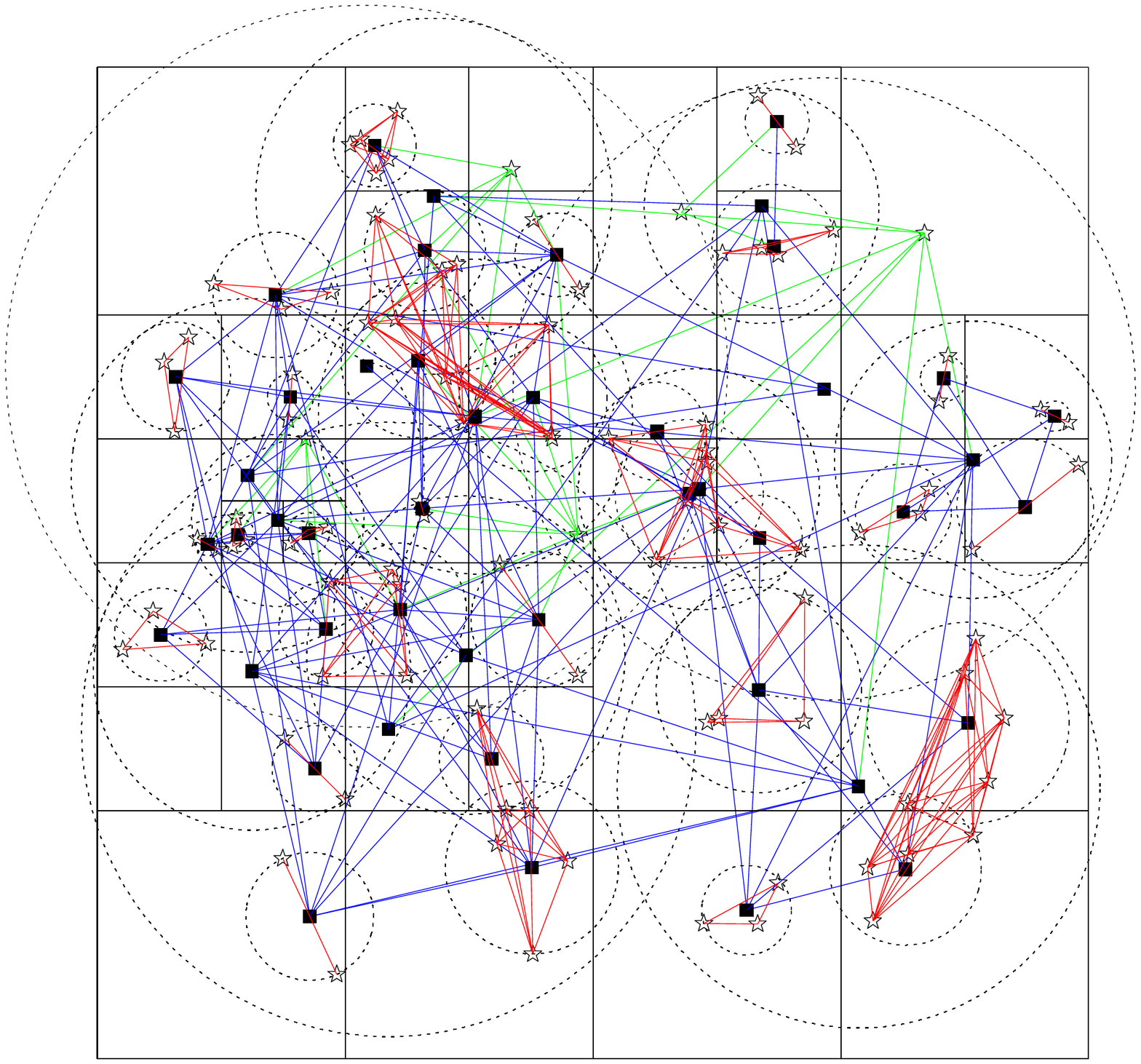}}
  }
  \caption{\small\label{fig:fmm}
    \textbf{Left:} Illustration of the geometry for the Taylor expansion used
    with the fast multipole method.
    \textbf{Right:} approximate calculation of the force for the same 100
    particles as in Fig.~\ref{fig:treecode} using the FMM, requiring 132
    cell-cell (blue), 29 cell-particle (green), and 182 particle-particle (red)
    interactions ($\theta=1$ and $n_{\max}=1$).
    }
\end{figure}
%

\vspace{-4mm}
\paragraph{The fast multipole method}
(FMM) also works with localised particle groups and for stellar systems is best
implemented using a tree structure \citep{Dehnen2000:falcON,Dehnen2002},
although the original proposal used a fixed grid \citep{GreengardRokhlin1987}.
In addition to expanding the Greens function $\phi(\vec{x}_b-\vec{x}_a)$ at the
\emph{source} positions $\vec{x}_a$ (as for the tree code), it also expands it
at the \emph{sink} positions $\vec{x}_b$ (with $\vec{z}_{\!B}$ the centre of the
group $B$ of sink particles, see also the left panel of Fig.~\ref{fig:fmm}):
\begin{equation}
  \label{eq:fmm:greens}
  \phi(\vec{x}_b-\vec{x}_a) \approx 
  \sum_{|\muin{n}|\le p}\, \sum_{|\muin{m}|\le p-n}
  \frac{(-1)^{\mathsf{n}}}{\muin{n}!\muin{m}!}\,
  (\vec{x}_b-\vec{z}_{\!B})^\muin{n}\,
  (\vec{x}_a-\vec{z}_{\!A})^\muin{m}\,
  \vec{\nabla}^{\muin{n}+\muin{m}}\phi(\vec{z}_{\!B}-\vec{z}_{\!A}).
\end{equation}
Inserting this into~(\ref{eq:est:pot}), yields for the potential generated by
all particles in $A$ and at any position $\vec{x}_b$ within $B$
\begin{eqnarray}
  \label{eq:pot:fmm}
  \hat{\Phi}_{A\to B}(\vec{x}_b) &\approx&
  \sum_{|\muin{n}|\le p}\frac{1}{\muin{n}!}
  (\vec{x}_b-\vec{z}_{\!B})^{\muin{n}} F_{\muin{n}}(\vec{z}_{\!B}),
  \\ \label{eq:ft}
  F_{\muin{n}}(\vec{z}_{\!B}) &=& \sum_{|\muin{m}|\le p-n}
  M_{\muin{m}}(\vec{z}_{\!A})\, D_{\muin{n}+\muin{m}}(\vec{z}_{\!B}-\vec{z}_{\!A})
\end{eqnarray}
with the multipole moments $M_{\muin{m}}(\vec{z}_{\!A})$ defined in
equation~(\ref{eq:mm}). The \emph{field tensors} $F_{\muin{n}}(\vec{z}_{\!B})$
are the coefficients of the Taylor series~(\ref{eq:pot:fmm}) for the potential
around $\vec{z}_{\!B}$. This dual expansion `at both ends' of all interactions
considerably speeds up the simultaneous computation of gravity for \emph{all}
particles, but brings no advantage over the tree code when computing the force
at a single position. The FMM algorithm works these equations backwards,
starting with an upward pass (as for the tree code) to compute the multipole
moments $M_{\muin{m}}(\vec{z}_{\!A})$ and widths $w_{\!A}$ for all cells.

The second part is the interaction phase, when the field tensors are evaluated
for each cell. This is achieved by the following algorithm starting with the
root-root interaction\footnote{Here, we describe the \textsf{falcON} algorithm
  \citep{Dehnen2002}, but essentially any multipole-based method which expands
  the Greens functions both at the source and sink positions and therefore
  performs cell-cell interactions qualifies as FMM.}. If for a mutual cell-cell
interaction $\theta (w_{\!A}+w_{\!B})<r=|\vec{z}_A-\vec{z}_B|$, then the field
tensors for the interactions $A\to B$ \emph{and} $B\to A$ are computed according
to equation~(\ref{eq:ft}) and added to $F_{\muin{n}}(\vec{z}_{\!B})$ \emph{and}
$F_{\muin{n}}(\vec{z}_{\!A})$, respectively. Otherwise, the interaction is split
into up to 8 new interactions by opening the bigger (in terms of $w$) of the two
cells. A cell self-interaction (like the initial root-root interaction) is
performed by simple direct summation if the cell contains only a few particles,
otherwise it is split into up to 36 new interactions.

Finally, in a down-ward pass the contributions from the parent cells are added
to the field-tensors of their daughters after applying the shifting formula
\begin{equation}
  F_{\muin{n}}(\vec{z}+\vec{x}) = \sum_{|\muin{k}|\le p-n}
  \frac{\vec{x}^{\muin{k}}}{\muin{k}!}\,F_{\muin{n}+\muin{k}}(\vec{z}),
\end{equation}
followed by the evaluation of gravity via equation~(\ref{eq:pot:fmm}) at the
sink positions within leaf cells. The total computational costs of this
algorithm are dominated by the interaction phase, which only requires
$\mathcal{O}(N)$ interactions for the computation of all $N$ particle forces.
This represents a substantial reduction from the $\mathcal{O}(N^2)$ for direct
summation. It is also a factor $\gtrsim10$ faster than the tree code for typical
$N\gtrsim10^6$.

\begin{figure*}
  \centerline{
    \resizebox{58mm}{!}{\includegraphics{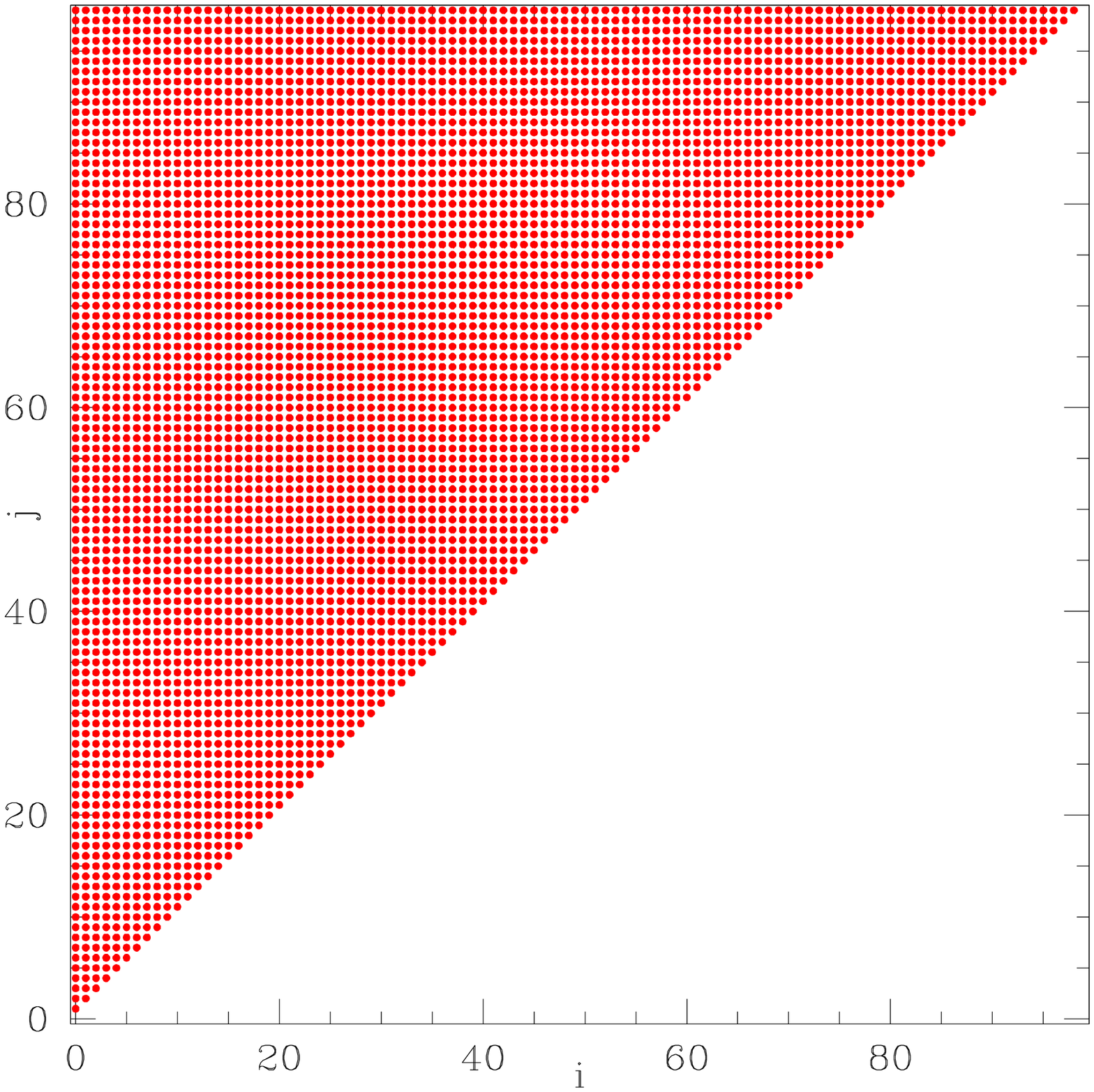}}\hfill
    \resizebox{58mm}{!}{\includegraphics{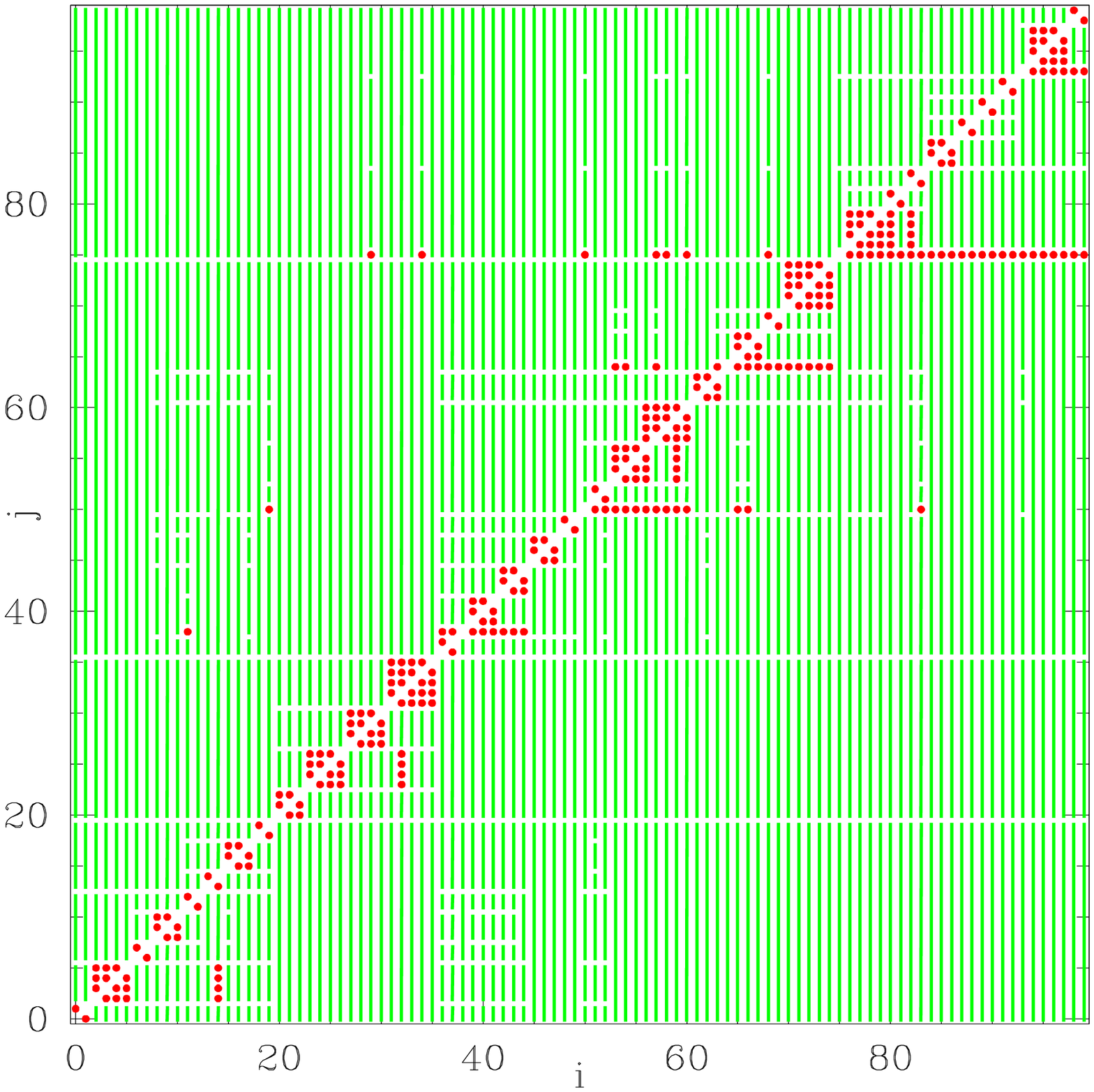}}\hfill
    \resizebox{58mm}{!}{\includegraphics{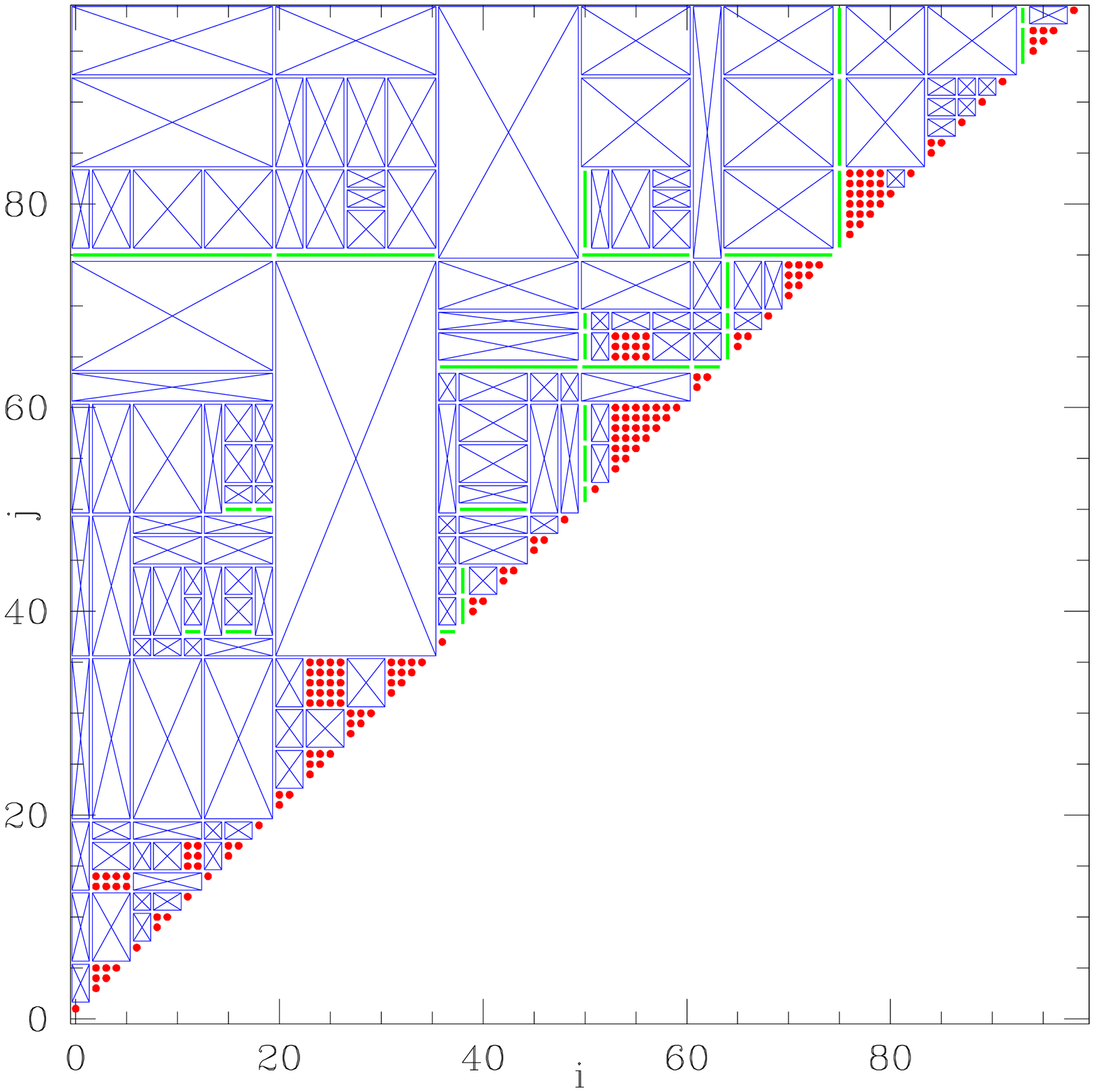}}
  }
  \caption{\small\label{fig:iact} 
    Logical interaction plots for the computation of all mutual forces between
    the same $N=100$ particles as in Fig.~\ref{fig:treecode}.
    \textbf{Left}: direct summation: every red dot corresponds to the
    force computation for one of $N(N-1)/2$ particle pairs.
    \textbf{Middle}: tree code: the approximated force for a particle
    $i$ from a cell containing many $j$ is represented by a green line
    (particles are ordered in their tree order: all particles within a tree cell
    are contiguous). Forces between close neighbours often cannot be
    approximated, resulting in the clustering of red dots along the diagonal.
    \textbf{Right}: using the FMM code \textsf{falcON}: each blue box
    corresponds to the approximation for the force between two cells. Unlike
    with the tree code, green lines represent \emph{mutual} interactions.
  }
\end{figure*}
%
\vspace{-4mm}
\paragraph{Some practical considerations}
\begin{enumerate}\itemsep1mm
\item The simple convergence criterion $|\vec{r}|<\theta w$ commonly used with
  the tree code and the FMM is merely geometric and therefore controls the
  \emph{relative} error for each interaction \citep{Dehnen2002}
  \begin{equation}
    \frac{|\delta\vec{\nabla}\Phi_{A\to B}|}{|\vec{\nabla}\Phi_{A\to B}|}
    \le \frac{(p+1)\theta^p}{(1-\theta)^2}.
  \end{equation}
  However, usually the forces from small (and hence nearby) cells are smaller
  than from bigger distant cells such that the \emph{absolute} force error is
  dominated by the few interactions with big cells. A better approach is
  therefore to try to \emph{balance} the force errors by a non-geometric opening
  criterion, for example using a mass-dependent opening angle. With this, one
  can obtain a total cost $<\mathcal{O}(N)$ at fixed approximation error
  \citep{Dehnen2002}.
\item For the un-softened Greens function $\phi=1/r$, the multipole expansion is
  reduced from three to two-dimensional indices when using spherical
  harmonics. Essentially, this exploits the fact that
  $D_{\muin{n}+2\hat{\muin{x}}} +D_{\muin{n}+2\hat{\muin{y}}}
  +D_{\muin{n}+2\hat{\muin{z}}}=0$ for any $\muin{n}$ (as a consequence of
  $0=\vec{\nabla}^2\phi$), such that the number of independent terms is reduced
  from $\binom{p+3}{p}$ to $(p+1)^2$. This reduces the costs for the FMM
  interaction operation~(\ref{eq:ft}) from $\mathcal{O}(p^6)$ to
  $\mathcal{O}(p^4)$, which may be further reduced to $\mathcal{O}(p^3)$ (by
  fast rotation methods, e.g.~\citep{PinchonHoggan2007}), $\mathcal{O}(p^2\ln
  p)$ (by fast Fourier methods \citep{ElliottBoard1996}), or even
  $\mathcal{O}(p^2)$ (using plane wave approximations
  \citep{ChengGreengardRokhlin1999}).
\item Both the distribution of force errors and the total computational costs
  depend on the opening criterion, the expansion order, and other numerical
  details. For an optimal choice of these numerical parameters, the
  computational cost depends non-trivially on the required approximation
  error. With collisionless $N$-body codes one usually requires only a modest
  relative force accuracy of $\sim10^{-3}$, in which case low-order techniques
  are sufficient.
\item Fig.~\ref{fig:iact} gives an alternative graphical comparison of direct
  summation, the tree code, and FMM for the computation of all $N$ forces. With
  direct summation, each pair-wise interaction is considered only once because
  the mutual forces satisfy $\vec{F}_{\!\!i\!j}= -\vec{F}_{\!\!j\!i}$, according
  to Newton's third law.

  \hspace{5mm}
  With the tree code, this natural symmetry between sinks and sources is broken:
  each interaction is one-sided. As a consequence, the full interaction matrix
  (except for the diagonal) has to be approximated and Newton's third law is
  violated, implying the loss of total-momentum conservation.

  \hspace{5mm}
  With the FMM on the other hand, each interaction is mutual, exploiting the
  natural symmetry of the problem and satisfying Newton's third law for each
  particle pair (though the approximated $\vec{F}_{\!\!i\!j}$ is not exactly
  aligned with the separation vector $\vec{x}_i-\vec{x}_{\!j}$). Only the upper
  half of the interaction matrix needs to be approximated. Further away from the
  diagonal, a single cell-cell interaction (blue box) approximates ever more
  particle-particle interactions.
  
\item From Fig.~\ref{fig:iact} one can also see that for the tree code the
  workload can be easily split between many processors. This is more difficult
  with direct summation or the FMM. For example, when distributing the
  particles amongst processors, it is not obvious which computes the forces
  between particles (or cells) residing on different processors (but
  see~\citep{2009MNRAS.398L..21S}).
\end{enumerate}

\vspace{-4mm}
\paragraph{Critique}
Apart from the violation of Newton's third law with the tree code (but not the
FMM), the only potential short-coming is a discontinuity of the forces: there
are some magic boundaries in space on either side of which a different
approximation will be used, depending on the opening criterion. As a
consequence, the approximated forces are not conservative across these
boundaries and the total energy of the $N$-body system is not exactly conserved
but exhibits some fluctuations, even in the limit of zero time step. Of course,
the amplitude of this effect can be reduced by decreasing the opening angle
$\theta$.

It appears that this issue has little if any effect on the validity of the
$N$-body model (there are no obvious differences between results obtained
using the tree code and methods which do not suffer from this problem). In
principle it should be possible to design a tree code or FMM without this
problem by smoothly interpolating between the force approximations on either
side of the magic boundaries. The natural adaptivity and efficiency of the
tree code and FMM make them versatile and powerful force solvers for
general-purpose $N$-body codes.
\subsubsection{Grid-based methods} \label{sec:force:grid}
Instead of solving the integral form~(\ref{eq:Phi}) or (\ref{eq:Phi:cosmo}) of
the Poisson equation, as with direct summation and its approximations,
grid-based methods solve its differential form
\begin{equation} \label{eq:Poisson:diff}
  \vec{\nabla}^2\Phi(\vec{x}) = 4\pi G\rho(\vec{x})
\end{equation}
after discretisation on a grid.

\vspace{-4mm}
\paragraph{Fast-Fourier-transform-based methods} achieve this in the Fourier
domain where the Poisson equation~(\ref{eq:Poisson:diff}) becomes $\vec{k}^2
\Phi(\vec{k}) =4\pi G\rho(\vec{k})$. The method first estimates the density on
the vertices of an equidistant grid by a technique which spreads the mass of
each particle across neighbouring cells (see~\citep{HockneyEastwood1988} for
details). Next, the Poisson equation is solved, using the fast Fourier
transform (FFT), for the value of the potential $\Phi$ on each grid vertex;
and finally the potential and force for each particle is found by
interpolation between vertices.

This method is fast (the cost for the FFT and the interpolation are
$\mathcal{O}(n_{\mathrm{grid}}\ln n_{\mathrm{grid}})$ and $\mathcal{O}(N)$,
respectively), but not effective for inhomogeneous particle distributions
typical for most $N$-body simulations, when in order to resolve the high-density
regions $n_{\mathrm{grid}}^3\gg N$ is required. 

The Fourier approach implicitly assumes a periodic tiling of all space with the
computational domain, such that it actually (approximately) solves
\begin{equation} \label{eq:Poisson:cosmo}
  \vec{\nabla}^2\Phi(\vec{x}) = 4\pi G \sum_{\muin{n}}\rho(\vec{x}+\muin{n}L).
\end{equation}
This corresponds not to equation~(\ref{eq:Phi}) but~(\ref{eq:Phi:cosmo}),
exactly as desired for cosmological $N$-body simulations of large-scale
structure formation. Therefore, FFT-based methods (or hybrid methods using FFT,
see below) are predominantly used with such simulations.

The periodic boundary conditions can be avoided either by simply doubling the
computational domain in each dimension or better by \citeauthor{James1977}'
\citep{James1977} method, which subtracts the contributions from the periodic
replicas of the computational box via a Fourier technique involving surface
charges. Combining this with a set of nested grids of increasing resolution
enables an efficient FFT-based force solver for inhomogeneous single stellar
systems, such as galaxies \citep{Magorrian2007}.

\vspace{-4mm}
\paragraph{Multi-grid techniques}
were pioneered in the west by \citet{Brandt1977} in 1977; they also interpolate
the density and potential between grid and particles, but solve the discretised
form of the Poisson equation, i.e.\ a large but sparse matrix equation, using
relaxation methods, such as Gauss-Seidel iteration. The basic idea exploits the
fact that on a coarser grid relaxation occurs faster because information travels
faster. The distribution of errors (the difference between the actual density
and that obtained via the discretisation of $\vec{\nabla}^2\Phi$ from the
current estimate for the potential) is first smoothed on the finest grid by a
few Gauss-Seidel iterations. After transferring the problem to a coarser grid,
the process is repeated on coarser and coarser grids until, on the coarsest grid
convergence is achieved. Then the problem is transferred back to finer and finer
grids, each time iterating until convergence.

Since the costs of an iteration shrinks by a factor eight when going to the
next coarser grid, the total cost is essentially determined by the costs of a
fixed number of iterations on the finest grid, and hence
$\mathcal{O}(n_{\mathrm{grid}})$ in theory
\citep{TrottenbergOosterleeSchuller2001}. However, as far as we are aware, in
practical applications for $N$-body simulations, the method is not
significantly more efficient than others.

The advantage of this technique over the FFT approach is that the grid does not
need to be equidistant, but can be locally adapted according to the particle
density. In fact, the structure of such an adaptively refined mesh is identical
to that of a shallow oct-tree, as used with the tree-code and FMM.

\vspace{-4mm}
\paragraph{Critique} Gravitational softening is implicit with grid methods, in
contrast to direct summation and its approximations, and depends on the grid
size. This means that for adaptive meshes the softening may vary along a
particle orbit, depending on the local mesh resolution. Since softening
unavoidably leads to a reduction in the estimated binding energy of a particle
(\S\ref{sec:soft:est}), these variations introduce an unphysical fluctuation
of particle binding energies. In non-equilibrium simulations, for example of
gravitational collapse, this leads to violation of energy conservation and
artificial secular evolution. This problem also occurs with individually
adapted softening lengths with explicit softening (as with the tree code or
FMM), but there a solution is known, see \S\ref{sec:adapt:eps}.

With adaptive multi-grid methods, it is natural to use different time steps
(within the block-step scheme) for particles living on different refinement
levels. Such a scheme can be substantially accelerated by advancing the
particles asynchronously: the particles on the coarser grid remain fixed while
those on the finer grid move and their gravity is approximated by using the
coarser-grid potential as boundary condition. However, such a method is neither
Hamiltonian nor time symmetric, potentially resulting in artificial secular
evolution (and violation of energy conservation).

\subsubsection{Basis function methods}
The idea of basis-function methods, pioneered by \citet{Clutton-Brock1972} in
1972 and later dubbed `self-consistent field code'
\citep{HernquistOstriker1992}, is to expand the mass density into basis
functions $\rho_{\muin{n}}(\vec{x})$ with coefficients $C_{\muin{n}}$
\begin{equation} \label{eq:scf:rho}
  \hat{\rho}(\vec{x}) = \sum_{\muin{n}} C_{\muin{n}}\,\rho_{\muin{n}}(\vec{x}),
\end{equation}
where $\muin{n}=(n,l,m)$ is a three-dimensional set of indices, and estimate the
potential as
\begin{equation} \label{eq:scf:Phi}
  \hat{\Phi}(\vec{x}) = -G\sum_{\muin{n}} C_{\muin{n}}\,\psi_{\muin{n}}(\vec{x}),
\end{equation}
where $4\pi\rho_{\muin{n}}(\vec{x})=-\vec{\nabla}^2 \psi_{\muin{n}}
(\vec{x})$. Usually, the sets of basis functions used are complete and
bi-orthogonal, i.e.\ satisfy
\begin{eqnarray}
  \label{eq:scf:ortho} \delta_{\muin{n}\muin{n}^\prime} &=&
  \int\!\mathrm{d}\vec{x}\,\rho_{\muin{n}}(\vec{x})\,
  \psi_{\muin{n}^\prime}(\vec{x}), \\
  \label{eq:scf:complete:rho}
  \delta(\vec{x}-\vec{x}^\prime) &=& \sum_{\muin{n}}
  \rho_{\muin{n}}(\vec{x}) \,\rho_{\muin{n}}(\vec{x}^\prime),
  \\
  \label{eq:scf:complete:psi} 
  \frac{1}{|\vec{x}-\vec{x}^\prime|} &=& \sum_{\muin{n}}
  \psi_{\muin{n}}(\vec{x}) \,\psi_{\muin{n}}(\vec{x}^\prime),
\end{eqnarray}
where the integral is over all space and the sums include all terms
$|\muin{n}|\to\infty$. Applying the bi-orthogonality relation
(\ref{eq:scf:ortho}) to the density estimate (\ref{eq:scf:rho}) gives
$C_{\muin{n}}=\int\!\mathrm{d}\vec{x}\,
\hat{\rho}(\vec{x})\,\psi_{\muin{n}}(\vec{x})$, which upon inserting of the
Monte-Carlo estimator for the mass density (equation~\ref{eq:est:mom} with $g$ a
delta-spike) yields
\begin{equation} \label{eq:scf:C}
  C_{\muin{n}}= \sum_i \mu_i\,\psi_{\muin{n}}(\vec{x}_i).
\end{equation}
Thus, the basis-function force solver first calculates the coefficients
$C_{\muin{n}}$ via equation (\ref{eq:scf:C}) and subsequently computes potential
and force at any position via equation (\ref{eq:scf:Phi}) and its
derivative. Since both of these operations are trivially split between multiple
processes and require minimal communication, this method is ideal for
computational parallelism.

Of course, in practice one has to truncate the expansion at some finite order
$\muin{n}_{\max}$. Ideally, the error made by this truncation is small, such
that the signal in the neglected coefficients $C_{\muin{n}>\muin{n}_{\max}}$ is
negligible. This is usually achieved by choosing the parameters of a given
basis-function set such that $\rho_{\muin{0}}$ already closely matches the
system modelled and higher-order terms merely describe deviations. A more
systematic approach is to \emph{design} the basis functions to match the
system at hand \citep{Weinberg1999} but also to truncate the expansion
smoothly according to the estimated signal-to-noise in the neglected
coefficients \citep{Weinberg1996}.

\vspace{-4mm}
\paragraph{Critique}
Of course, the basis-function approach is not suitable for modelling wildly
dynamic situations, such as galaxy mergers, but only for near-equilibrium
dynamics, when the stellar system deviates only slightly from the smooth
zeroth-order basis function. An advantage of the method is the effective
softening, which usually is small in high-density regions where the the basis
functions vary mostly. Unlike Greens-function softening, the method does not
bias the force of a density cusp ($\rho\propto r^{-\gamma}$) \emph{if} the basis
functions contain the same power-law cusp \emph{and} the $N$-body system is
centred on the origin of the expansion. In fact, off-centring is a significant
problem with this method.

It has been argued that the basis-function method is ideally suited for such
near-equilibrium systems, because the computational cost scales linear with
$N$. However, this view is too simplistic, since it is futile to increase $N$
but not also the force resolution, i.e.\ $\muin{n}_{\max}$. Simple arguments
based on equivalence to Greens-function softening suggest that the optimum
resolution for given $N$ requires an increase of $\muin{n}_{\max}$ such that the
overall computational cost scale like $\mathcal{O}(N^2)$, as for straightforward
direct summation \citep{Dehnen2001}.

One may hope to alleviate this problem by a continually adapting the
lowest-order basis function \citep{Weinberg1999} such that $\muin{n}_{\max}$ can
be kept low. However, this approach changes the approximated Greens function
\begin{equation}
  \phi(\vec{x}-\vec{x}^\prime) = \sum^{\muin{n}_{\max}}_{\muin{n}}
  \psi_{\muin{n}}(\vec{x})\,\psi_{\muin{n}}(\vec{x}^\prime)
\end{equation}
during the simulation and therefore introduces an artificial time dependence
into the approximate Hamiltonian. At best, this only destroys energy
conservation, but more likely has other adverse effects which are less obvious
to identify.

The basis-function method may be most useful for simulations with constrained
symmetry. For example, enforcing spherical symmetry simply amounts to setting
$C_{nlm}\propto\delta_{l0}\delta_{m0}$ when using a basis based on
spherical-harmonics. This reduces the computational costs substantially, even
when including many radial basis functions, thus enabling extremely fast
spherically symmetric simulations with large $N$. The basis-function method is
certainly very useful as force solver for other purposes, for example to
approximate the density and/or potential of an external system or to model the
potential in perturbation analyses.

\subsubsection{Hybrid methods}
The advantages and disadvantages of the various force solvers naturally lead to
the concept of hybrid methods, to avoid the respective disadvantages. Most
relevant in this context is presumably the usage of FFT-based methods to obtain
periodic boundary conditions, which are desired in cosmological simulations. The
disadvantage of the FFT-based method (also called `PM' particle-mesh) is the
lack of resolution and adaptivity on small scales. In the 1980s, the early days
of cosmological simulations, the combination of PM with direct summation (`PP':
particle-particle) for the computation of the near-neighbour force deficit (the
difference between the average FFT force and that for the desired softening
length) enjoyed some popularity as `P$^3$M' codes
\citep{EfstathiouEtAl1985}. Later, this approach has been improved by replacing
direct summation with a tree code (`TreePM' \citep{BodeOstriker2003}), but we
are not aware of the obvious FMM-PM combination. Combinations of the multi-grid
method with the FFT are straightforward (the FFT is used as force solver on the
coarsest grid \citep{KravtsovKlypinKhokhlov1997}).

Other hybrid methods combine the basis-function approach in a subset of the
spatial dimensions with a grid in the remaining, for example using spherical
harmonics with a radial grid \citep{McGlynn1984, TrentiBertinVanAlbada2005}.

\subsection{Recent numerical developments and challenges}
\label{sec:less:recent}
Collisionless $N$-body simulations have benefitted enormously from the
incredible increase in computer power, in particular as the computational costs
only increase like $N\ln N$ (for the tree code and grid methods). This combined
with massive parallelisation on many thousand cores has allowed very large $N$
simulations. A further increase by a factor $\sim10$ should be possible just by
using the FMM as force solver (though implementing the FMM efficiently in
parallel is challenging). Another welcome numerical advancement would be better
time-stepping methods (see the discussion at the end of
\S\ref{sec:step:choice}).

However, the main challenge of contemporary applications of collisionless
$N$-body methods lie not in the method, but in the astrophysics it misses out:
any non-gravitational interaction. Baryonic matter contributes only 16\% of all
matter on large scales \citep{2011ApJS..192...18K} and inter-stellar gas
contributes only little to the mass of individual galaxies (in the Milky Way the
mass ratio between gas and stars is about 1:9). However, the gas not only
interacts gravitationally (with all other matter), but can directly dissipate
(and absorb) energy in form of radiation, and therefore behaves fundamentally
differently from point-mass particles.

One big challenge of contemporary astrophysics is to model the formation of
galaxies ab initio. To this end, many complex astrophysical processes and
phenomena must be modelled, such as the multi-phase nature of the gas, its
condensation to stars and active galactic nuclei (AGN), and their feedback (via
winds and radiation) onto the gas (all these process may be summarised as
`baryon physics'). While hydrodynamics is comparatively straightforward to add
to the method, e.g.\ via smoothed particle hydrodynamics \citep{Price2010}, most
of the baryonic physics is not. This is because these processes are themselves
not properly understood and can only be modelled via parametric `sub-resolution'
models. For example, a gas particle is turned into a star particle according to
some parametric model of our current understanding of the star formation
process. While substantial progress has been made, the challenges are still
formidable, not only because of our limited astrophysical understanding of the
baryon physics, but also since much higher numerical resolution may be required
for convergence than with simple gravity-only $N$-body experiments.

\subsection{Past, recent, and future astrophysical modelling}
\label{ref:less:astr}
As in \S\ref{sec:coll:astr}, we provide a very brief summary of selected (by
personal opinion) highlights of astrophysical results based on collisionless
$N$-body simulations (apology to anybody who feels missed-out).

One of the earliest results was the modelling by
\citeauthor{KozlovSyunyaevEneev1972} in 1972 [\citenum{KozlovSyunyaevEneev1972},
  see also \citenum{EneevKozlovSunyaev1973,
    1941ApJ....94..385H}]\footnote{Sadly, this excellent work is only
  little-known (63 citations), largely because publishing in the west was
  extremely difficult for Soviet scientists during the cold-war era. The later
  study by Toomre \& Toomre (ApJ {\bf 178}, 623) with essentially the same
  simulation set-up but only $N=120$ (compared to $N=2000$) is much more widely
  known (over 1700 citations).} of tidal interactions of rotationally supported
disc galaxies, demonstrating a large variety of observed phenomena, such as
tidal arms and bridges. Simulations of galaxy interactions could explain many
other observed phenomena such as shells, ripples, and other `fine structure'
\citep{Quinn1984, DuprazCombes1986, HernquistQuinn1988, HernquistQuinn1989}, as
well as ring galaxies \citep{LyndsToomre1976}. Also, the merger origin of
elliptical galaxies received strong support from $N$-body simulations
\citep{Barnes1988,Barnes1990}.

The bar instability of isolated disc galaxies was discovered in the first
simulations of such systems \citep{MillerPrendergastQuirk1970, Hohl1971},
providing a natural explanation of the high frequency of barred galaxies. More
recent simulations suggest a close connection between the dynamics of bars and
outer rings \citep{AthanassoulaRomeroGomezMasdemont2009} and demonstrate the
importance of a dark-matter halo as an angular-momentum absorber
\citep{Athanassoula2003,2009ApJ...697..293D}. The buckling instability of bars
was first seen in $N$-body simulations \citep{CombesSanders1981}, though the
identification of `boxy' or `peanut-shaped' bulges as side-on bars was only made
later. Simulations of disc galaxies also showed spontaneous formation of
transient and/or long-lived spiral features \citep{Zang1976PhD}, which prompted
the theoretical model of swing-amplification by Toomre \& Zang
\citep{Toomre1981}.

`Cosmological' simulations based on cold dark-matter match the observed
large-scale structure of the universe \citep{DavisEtAl1985,
  SpringelFrenkWhite2006}, while hot (e.g.\ neutrino) dark-matter is ruled out
\citep{WhiteFrenkDavis1983}. The dark-matter haloes formed in such simulations
possess a universal density profile
\citep{DubinskiCarlberg1991,NavarroFrenkWhite1996}, which varies between
$\rho\propto r^{-1}$ at $r\to0$ and $\rho\propto r^{-3}$ to $r^{-4}$ at large
radii, and triaxial shapes \citep{DubinskiCarlberg1991,WarrenEtAl1992}.  This is
now observationally confirmed on galaxy cluster scales
\citep{SahaReadWilliams2006, SahaRead2009, UmetsuEtAl2011}. Finally, such
simulations predict a wealth of substructure that should be present orbiting the
Milky Way, and in the Local Group that is not (yet) observed
\citep{1999ApJ...522...82K, 1999ApJ...524L..19M}.

Confronting such predictions, as well as those of almost all applications of
collisionless $N$-body simulations, with observations requires an extension of
the method to include baryonic physics (see \S\ref{sec:less:recent}). The goal
for the coming decades will be realistic simulations of galaxy formation and
evolution, which will allow us to test both our current cosmological model, and
galaxy formation theories.

\section{Validation} \label{sec:valid}
A very important issue with $N$-body techniques is the validation of the
numerical method in general and for the specific purpose in particular. While
a 100\% validation is never possible, there are several powerful validation
techniques available. In particular for collisionless $N$-body techniques, every
project requires a validation to ensure that the numerical method (which
includes the values of numerical parameters such as $N$) guarantees the desired
numerical accuracy, thus avoiding systematic errors driven by numerical
artifacts.

\vspace{-5mm}
\paragraph{Validation of the method}
Before even considering the application of a certain $N$-body program
(collisional or collisionless) for a particular modelling purpose, the values of
numerical parameters (controlling time step, force approximation, etc.)
required for correctness must be established. This can be done in two ways:
first by simulating simple but non-trivial situations for which the correct
outcome is known from other means (such a perturbation analysis or, in the case
of collisional $N$-body code by using any of the alternative methods of
\S\ref{sec:collalternative}); and second (for collisionless $N$-body methods) by
re-simulating a certain situation with ever better numerical accuracy, which
should converge to a (hopefully correct) answer.

\vspace{-5mm}
\paragraph{Validation by Monitoring}
One of the most straightforward validation requirements is the actual
numerical conservation of total energy, momentum, and angular momentum. While
with a finite time step energy conservation can never be fully achieved, a
relative energy conservation to 3-4 digits is usually considered sufficient in
practice for collisionless $N$-body techniques, while for collisional methods
5-6 digits appear to be the norm. A systematic trend of energy hints at an
artificial secular evolution and is much more problematic than mere
fluctuations. The same applies to momentum and angular momentum.

In order to monitor whether gravitational softening introduces significant
bias (if for instance many particles overlap within the softening length),
one can compare the virial $W$ and the potential energy $V$, defined as
\begin{equation} \label{eq:W:V}
  W = \sum_i\mu_i\;\vec{x}_i\cdot\vec{\nabla}
  \left[\hat{\Phi}_{\mathrm{self}}(\vec{x}_i) +
  \Phi_{\mathrm{ext}}(\vec{x}_i)\right],
  \qquad
  V = \sum_i\mu_i\,
  \left[\tfrac{1}{2} \hat{\Phi}_{\mathrm{self}}(\vec{x}_i) +
  \Phi_{\mathrm{ext}}(\vec{x}_i)\right]
\end{equation}
with $\hat{\Phi}_{\mathrm{self}}$ and $\Phi_{\mathrm{ext}}$ denoting,
respectively, the (approximated and softened) $N$-body potential and any
external potential. For pure Newtonian gravity (unsoftened), one has $W=V$ and
deviations from this equality are indicative of the global importance of the
gravity reduction due to softening.

\vspace{-5mm}
\paragraph{Validation by simulation}
When the effect of some particular perturbation is to be modelled, for example
of an infalling satellite, it is important to ensure that, at the resolution at
which the simulation results will be interpreted, a control simulation without
the perturbation behaves as expected. Sometimes, this may not be sufficient, for
example when the perturber gives rise to additional astrophysical effects not
present in and hence not scrutinised by the control simulation. In such a
situation one must employ convergence tests: an equivalent simulation at
significantly higher resolution should give consistent results. Even this may
fail when the simulations converge to the wrong answer.

\vspace{-5mm}
\paragraph{Validation by comparison}
The only way to to shield against this problem is to re-simulate with a
completely different $N$-body method. For example, various $N$-body tests have
been done within the ESF-funded astro-sim project
(\texttt{http://www.astrosim.net/code}). With collisionless $N$-body methods
this is often the only reliable validation technique, since there are hardly any
non-linear problems with solutions known from other methods (against which one
could compare). Such projects are important and have demonstrated that
cosmological codes now agree at the 10\% level for important metrics
\citep{HeitmannEtAl2008}. This is excellent progress, but not good enough for
next generation cosmological probes like \textsf{Euclid}
(\texttt{http://sci.esa.int/euclid}).

\paragraph{Validation and interpretation}
Simulation results, in particular when unexpected, should never be trusted at
face value, but an effort must be made to \emph{understand} them in
astrophysical terms. This is very often non-trivial, because the highly
non-linear and complex dynamics is not compatible with any simplifying
assumptions that would allow insight by means of approximate analytical models.
However, unlike the situation with real stellar systems and galaxies, the
$N$-body model can be analysed (``observed'') in full six-dimensional
phase-space allowing a much better handle on the dynamical processes than nature
itself. Contemporary analysis techniques for $N$-body data are still somewhat
immature, but their discussion beyond the scope of this paper.

\section{Conclusion} \label{sec:conclusion}
We have reviewed some of the latest numerical techniques for modelling
collisional and collisionless $N$-body systems. Our focus throughout has been on
purely gravitational $N$-body simulations, with a view to presenting the key
numerical algorithms. Over the past $\sim 50$ years of $N$-body calculations,
the field has undergone dramatic changes. Improved software algorithms,
specialised hardware, and efficient parallel programming have meant that $N$ has
kept pace with Moore's Law, nearly doubling every two years. This has allowed us
to simulate the most massive star clusters, galaxies, and the Universe as a
whole with increasing precision.

It is clear that our modelling of gravity is in good shape. Different codes and
techniques give converged results, while both collisional and collisionless
simulations continue to push to larger $N$. However, the coming challenge will
be building believable models of baryonic processes in the Universe. This is
beyond the scope of this short review, but is the key challenge for future
$N$-body simulations on all scales.


\end{document}